\documentclass[prb,twocolumn,showpacs,preprintnumbers,amsmath,amssymb,showkeys]{revtex4}
\usepackage[T1]{fontenc}
\usepackage[latin9]{inputenc}
\usepackage{graphicx}
\usepackage{amssymb}
\newcommand{\beq}{\begin{equation}}
\newcommand{\eeq}{\end{equation}}
\newcommand{\ket}[1]{\mbox{$ \mid #1\, \rangle$}}
\newcommand{\bra}[1]{\mbox{$ \langle\, #1\mid$}}

\newcommand{\vecg}[1]{\mbox{${\boldsymbol #1}$}}

\newcommand{\vecb}[1]{\mbox{\bf\scriptsize #1}}

\begin{document}
\title{Electronic doping of graphene by deposited transition metal atoms}
\author{Jaime E. Santos$^{1,3}$, Nuno M. R. Peres$^{2,5}$, Jo\~{a}o M. B. Lopes dos Santos$^3$
and Ant\'{o}nio H. Castro Neto$^{4,5}$}

\affiliation{$^1$ Max Planck Institute for the Physics of Complex Systems,
N\"othnitzer Str. 38, D-01187 Dresden, Germany; 
$^2$ Centro de F\'{i}sica and Departamento de F\'{i}sica, Universidade do Minho,
P-4710-057 Braga, Portugal; 
$^3$ CFP and Departamento de F\'{i}sica, Faculdade de Ci\^{e}ncias, Universidade
do Porto, 4169-007 Porto, Portugal;
$^4$ Department of Physics, Boston University, 590 Commonwealth Avenue, Boston, 
Massachusetts 02215, USA;
$^5$ Graphene Research Centre, and Department of Physics, National University
of Singapore, 2 Science Drive 3, 117542, Singapore;
}

\email{jaime.santos@cpfs.mpg.de}

\date{today}
\begin{abstract}
We perform a phenomenological analysis of the problem of 
the electronic doping of a graphene sheet by deposited transition metal atoms,
which aggregate in clusters.
The sample is placed in a capacitor device such that the electronic doping
of graphene can be varied by the application of a gate voltage and such that
transport measurements can be performed via the application of a (much smaller)
voltage along the graphene sample, as reported in the work of
Pi et al. [Phys. Rev. B \textbf{80}, 075406 (2009)].
The analysis allows us to explain the thermodynamic properties
of the device, such as the level of doping of graphene and the ionisation potential
of the metal clusters in terms of the chemical interaction between graphene
and the clusters. We are also able, by modelling the metallic clusters as 
perfect conducting spheres, to determine the scattering potential due to these 
clusters on the electronic carriers of graphene and hence the contribution of these 
clusters to the resistivity of the sample. The model presented is able to
explain the measurements performed by Pi et al. on Pt-covered graphene samples
at the lowest metallic coverages measured and we also present a theoretical
argument based on the above model that explains why significant deviations
from such a theory are observed at higher levels of coverage.
\end{abstract}



\maketitle
\section{Introduction}
\label{secInt}
Graphene was discovered in late 2004 \cite{novo1,pnas}. This
material is a one-atom thick sheet of carbon atoms, arranged in a
honeycomb lattice. This structure is not a Bravais lattice and graphene
is described in terms of a triangular lattice with a two-atom basis.
A simple nearest-neighbour tight-binding approximation of the electronic
Hamiltonian in graphene reveals that such lattice structure leads
to a dispersion relation that is linear around two specific points
of the Brillouin zone. Since the Fermi level of graphene
lies at these points, its quasi-particles behave in a continuum approximation
as massless relativistic fermions with a speed of light equal to the
Fermi-velocity $\approx 10^{6}ms^{-1}$ (see \cite{rmp} and the recent
review \cite{tapash}).

The properties of graphene and its special geometry make it a very
interesting candidate for applications in nano-electronics. Recent
research has revealed other possible applications, in solar cell technology
\cite{solar}, in liquid crystal devices \cite{liquid}, in single
molecule sensors \cite{sensors}, and in the fabrication of nano-sized
prototype transistors \cite{dai}. 

\begin{figure}
\begin{center} 
\includegraphics[clip,width=5cm]{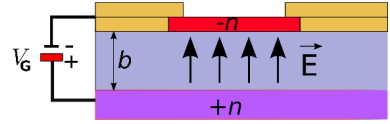}
\par
\end{center}
\caption{Capacitor device whose upper electrode is a single graphene sheet. The application of
a gate voltage imposes a certain level of electron or hole doping on graphene, continuously
increasing or lowering its Fermi level. The application of a potential difference
between the upper contacts allows for transport measurements to be performed.}
\label{fig:1} 
\end{figure}

Transport measurements on graphene devices \cite{novo1} have become
standard and can be performed under different doping conditions. Given
the location of the Fermi level and the absence of a band gap between
the valence and conduction bands in undoped graphene, one can continuously
control the level of doping simply by the application of a gate voltage
in a geometry where graphene acts as the upper (grounded) electrode
of a capacitor. The lower electrode is composed of
silicon, whereas the dielectric medium in between is SiO$_2$. Metal contacts placed
on top of the graphene sheet allow for the realisation of transport
measurements at different gate voltages, and hence at different levels
of doping, with great flexibility. The system has an overall
thickness of $b\approx 300$ nm (see figure \ref{fig:1}).

The measurement of the transport properties using such devices can be 
used to determine the influence of different physical effects on both
the AC and DC conductivities. One can investigate the influence of electron-electron 
interactions, of impurities, or of the presence of elastic ripples in the 
graphene sheet on the transport properties of this semi-metal \cite{rmp}.

It well known that undoped graphene, when analysed from the point of view 
of a self-consistent theory or the renormalisation group, presents a finite 
conductivity with an universal value of  $4e^2/(\pi h)$
\cite{schon,zheng,ostrovsky,katsC,peres06,stauber08a,dora,leconte,
lherbier100,lherbier106}, regardless of the scattering mechanism that limits 
conductivity in graphene. The experimental measurements\cite{geim07,tan07} 
point to a somewhat higher value for this quantity, equal to $4e^2/h$. This 
latter value is also obtained in studies of numerical diagonalisation of graphene's tight-binding 
Hamiltonian with add-atoms acting as the source of disorder\cite{lherbier101,ferreira}.
In the case of doped graphene, the behaviour of the conductivity markedly depends
on the scattering mechanism that limits such quantity. It is therefore
essential to clarify the nature of such a mechanism. The research community has held 
two opposing views, namely charged (Coulomb) or short-range scatterers \cite{kats}, but 
recent experiments \cite{pono} seem to show the latter mechanism as the prevailing one,
even if it is agreed that charged scatterers also play a role \cite{peresrmp}.

In this paper, we will consider the contribution to the conductivity of one
particular type of short-range disorder, namely that induced by the deposition
of transition metal (TM) atoms in graphene \cite{pi}. This type of disorder
is always present on devices such as those depicted in figure \ref{fig:1},
due to the diffusion of metallic atoms from the contacts into the graphene
sheet. The adsorption of graphene on TM surfaces has been extensively 
studied, both experimentally \cite{marchini,parga, sutter, preo}, 
as well as theoretically \cite{chan, mao, wang,jiang,gio1,gio2}. 
In TM surfaces for which the adsorption process (physisorption) 
preserves the conical nature of the graphene bands
close to the Dirac point (Al, Ag, Cu, Au, Pt), 
the authors of \cite{gio1,gio2} have shown that the levels of electron or hole
doping of graphene that they have found in their DFT studies can be explained by 
the relative value of the bulk work-functions of
graphene and of that of the transition metal to which graphene is adsorbed. However,
in order to explain the electron-doping of graphene
in cases where its work-function is lower than that of the transition metal (Ag, Cu),
the authors invoked the existence of a chemical interaction between graphene
and the underlying metal substrate, which plays a significant
role in the formation of surface dipoles\cite{seitz,herring,lang1,heine,lang2,jsilva}. 
The existence of such an interaction was confirmed in the experimental 
transport studies of \cite{pi}, performed on a graphene sheet where
TM atoms were deposited, which was part of device such as that of figure 
\ref{fig:1}. The authors of this study have found that in the 
case of low coverage of graphene by Pt, the metal with the highest work 
function studied theoretically by \cite{gio1,gio2}, graphene is also 
electronically doped by Pt, becoming hole-doped at higher coverages. 
The authors stated that the high levels of electron-doping that they have
found at low coverages were caused by an increased chemical interaction 
between graphene and the TM atoms, due to the short-distance (less than $3$ A) 
between the two species (this distance is equal to  $3.3$ A in the full coverage regime).
Furthermore, the AFM pictures obtained seem to show that the transition metal atoms 
aggregate in clusters at low coverage (see also \cite{McCreary}). It is nevertheless unclear 
whether the proximity between the clusters and graphene is sufficient to justify
the level of doping of graphene.

The purpose of this paper is threefold. Firstly, we wish to introduce a
framework that allow us to discuss the problem of charge doping of
graphene by transition-metal clusters with generality from a thermodynamic 
point of view. This framework  will be of a phenomenological nature and 
it will involve some simplifying assumptions, but it will already 
contain the main ingredients that will need to be considered in
a more fundamental approach. Secondly, the same type of phenomenological analysis
will be extended to the problem of electronic scattering in graphene caused by 
the presence of the said clusters. We will show, following \cite{fogler} that, 
despite the charged nature of the clusters, the scattering potential that 
they create is of a short-ranged nature (see also \cite{kats}). The domain of
validity of the semi-classical approximation of independent clusters that we
are using is also discussed. Thirdly, these two elements of the theory will be 
used to interpret the above experiments from a quantitative point of view. This application
of the theory will also serve to illustrate its overall limitations and
we will provide physical arguments that show why more elaborate approaches 
are needed. 

The structure of this paper is as follows: in section \ref{secO}, we will discuss
the doping of graphene by metal clusters in a capacitor device based on a phenomenological 
model that treats each cluster as a perfectly metallic object kept at a constant potential
dependent on the amount of charge in the cluster. The minimisation of the internal
energy of the system at $T=0$, subjected to overall charge conservation, will allows
us to obtain the equilibrium conditions that determine the level of doping of the
graphene sheet. One can show, for equally charged clusters, 
that the level of doping can be written in terms of the bulk work-functions 
of the different components of the system, of the gate voltage applied to the
device, and of a parameter that characterises the 
{\it effective} chemical interaction between the graphene sheet and each individual cluster. 
The numerical value  of this parameter is determined by two different
contributions: the first contribution is due to the induced surface dipole
of graphene and of the metallic clusters caused by the presence of the
other components of the system; the second contribution is the correction to the 
Fermi energy of a cluster due to its finite size. Specialising to the 
case of spherical clusters, we can estimate the magnitude of the chemical interaction 
using the measured values by\cite{pi} of the gate voltage that is necessary
to apply to Pt and Ti-covered\cite{Tichan} graphene samples to bring graphene 
to an uncharged state, where the conductivity is a minimum. These values are 
dependent on the concentration of metallic atoms per unit cell as well as on 
which metallic element is deposited on graphene. In section
\ref{secB}, we will consider the form of the scattering potential created by a spherical cluster,
following the model of \cite{fogler} and its contribution to the resistivity
of the sample, within the First Born Approximation (FBA) and we will 
compare our results with the transport measurements of\cite{pi}, performed
on Pt-covered graphene samples at low coverage. We will also show why the
theory presented is not adequate to explain the measurements performed at higher
coverages, both for Pt and Ti-covered samples. We will determine, using
the above model, the linear dimension of the region in graphene where the
charge donated by the cluster to this material is contained and 
show that, except for the lowest coverages considered in the experiments 
of\cite{pi}, this quantity is comparable to the average distance between
clusters, even for heavily-doped graphene. In section \ref{secC}, we will present 
our conclusions. In appendix \ref{appO}, we will derive an expression for the 
cluster's ionisation potential that will be used in the main text, based on the same 
thermodynamic arguments that were used in section \ref{secO}.
Finally, in appendix \ref{secA}, we will derive, using the method of images, 
the electrostatic contribution to the ionisation potential of a single spherical cluster,
a result that will be shown to be in agreement with that of appendix \ref{appO}.
This derivation will also allow us to obtain the capacitance of the system 
composed of the metallic cluster and of the graphene plane, as well as the 
electrostatic potential due to a charged spherical cluster close to a 
grounded plane, a quantity that enters in the calculations performed in section \ref{secB}.

\section{Electronic doping of graphene by deposited metal clusters}
\label{secO}

At $T=0$, the internal energy of a composite system of $k$ conductors can be 
written\cite{hohenberg,kohn}, assuming that the electrons and the
(immobile) ions of the different species interact with each other via 
the bare Coulomb interaction (i.e. one is including the contribution of the low lying
electronic orbitals explicitly in the energy), as
\begin{eqnarray}
{\cal E}&=&G(N_1,\ldots,N_{k})\nonumber\\
&&\mbox{}+\frac{1}{2}\sum_{i,j}\int_{V_i}\,d^dr\,\int_{V_j}\,d^dr'\,
\frac{\rho^P_i(\vecg{r})\,\rho^P_j(\vecg{r}')}{4\pi \epsilon_0\mid\vecg{r}-\vecg{r}'\mid}\,,
\label{eq0m1}
\end{eqnarray}
where the functional $G(N_1,\cdots,N_{k})$ includes the 
kinetic, exchange and correlation energies of the electrons and where the
second term includes the effect of the ion potential on the electrons
and the Hartree energy of these electrons, 
with $\rho^P_i(\vecg{r})=\rho^e_i(\vecg{r})-\rho^{ions}_i(\vecg{r})$ being
the plasma charge density. The indices $i, j$ run over $1,\ldots,k$.
One can write $\rho^P_i(\vecg{r})=\rho^n_i(\vecg{r})+\delta\rho_i(\vecg{r})$, where
$\rho^n_i(\vecg{r})$ is the charge density in the neutral ground state of each
conductor and
$\delta\rho_i(\vecg{r})$ is the excess charge density of that conductor due to
charge exchange with the others. In particular,
$\int_{V_i}\,d^dr\,\delta\rho_i(\vecg{r})=Q_i$, the total unbalanced
charged contained in conductor $i$. Using this decomposition, one can write
the ground-state energy, up to a constant term, as 
\begin{eqnarray}
{\cal E}&=&G(N_1,\ldots,N_{k})+\sum_i\,\int_{V_i}\,d^dr\,
{\cal V}_i^n(\vecg{r})\,\delta\rho_i(\vecg{r})\nonumber\\
&&\mbox{}+\frac{1}{2}\sum_{i,j}\int_{V_i}\,d^dr\,\int_{V_j}\,d^dr'\,
\frac{\delta\rho_i(\vecg{r})\,\delta\rho_j(\vecg{r}')}{4\pi \epsilon_0\mid\vecg{r}-\vecg{r}'\mid}\,,
\label{eq0m2}
\end{eqnarray}
where ${\cal V}_i^n(\vecg{r})$ is the potential on conductor $i$ due
to itself and the other conductors, each in a neutral state. In a classical
approximation, $\delta\rho_i(\vecg{r})$ will be non-zero only close
to the surface of the conductors, and we can write the above expression
in a capacitor approximation:
\begin{eqnarray}
{\cal E}&=&G(N_1,\ldots,N_{k})+\sum_i\,D_i\,Q_i\nonumber\\
&&\mbox{}+\frac{1}{2}\sum_{i,j}\,C_{ij}^{-1}\,Q_i\,Q_j\,,
\label{eq0m3}
\end{eqnarray}
where $C_{ij}^{-1}$ is the inverse cross-capacitance between conductors
$i$ and $j$ and $D_i=\overline{{\cal V}_i^n(\vecg{r})}$ is the surface dipole
of conductor $i$\cite{lang2}, which is the average value of the
electrostatic potential within that conductor. Note that the surface 
dipole of a conductor is computed with the remaining conductors present, 
but in a neutral state. Thus, one expects that such surface dipoles will 
depend both on the geometry of each conductor and also on the 
presence of the other conductors if the distances between them are on the 
atomic scale.  

\begin{figure}
\begin{center} 
\includegraphics[clip,width=7cm]{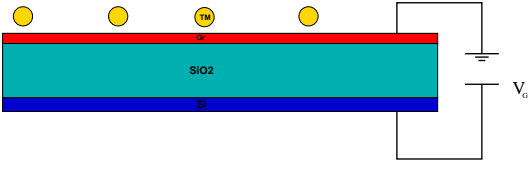}
\par
\end{center}
\caption{Schematic representation of model that is considered in this paper.
Transport measurements can be performed at different coverages as more atoms
are deposited by MBE.}
\label{fig:2} 
\end{figure}

As stated above, the experiments of reference \cite{pi} were performed 
on a devices similar to that depicted in figure \ref{fig:1}, with the
transition metal atoms deposited by molecular beam epitaxy (MBE)
on the graphene sheet at different
coverages $c_S=N_{am}/n_u$ of metallic atoms per unit cell of graphene,
where $N_{am}$ is the total number of deposited metallic atoms and
$n_u$ is the number of graphene's unit cells. In order to model such a device, we 
will assume that the atoms aggregate in $N_c$ identical clusters, which are
randomly distributed above the area 
$A_g=n_u A_c$ of the graphene sheet, where $A_c$ is the area of the graphene
unit cell. We also assume that these clusters are all equally charged.
The graphene sheet is kept at zero potential. At a distance $b\approx300$ nm below it,
one places a Si layer, with the space in between filled with SiO$_2$,
a medium of permittivity $\epsilon=3.9\,\epsilon_0$.
The graphene sheet and the Si layer are connected to a battery such that a constant gate-voltage 
$V_G$ is kept between them ('plane-capacitor model', see figure \ref{fig:2}).
A single cluster-graphene subsystem is assumed to possess a joint capacitance $C_S$
(which is computed for spherical clusters in appendix \ref{secA}).
The capacitance of the graphene/SiO$_2$/Si device is given by $C_{Si}=\epsilon A_g/b$. 
We assume that the cross-capacitance effects between different clusters
are only due to the presence of the grounded graphene plane, an assumption
that is correct for small $c_S$\cite{NoteCC}.
With such assumptions, the internal energy (\ref{eq0m3}) can be written
for this system, as
\begin{eqnarray}
{\cal E}&=&G(N_S,\ldots,N_S,N_g,N_{Si})-e\,N_c\,D_S(N_S-N_S^0)\nonumber\\
&&\mbox{}-e\,D_g\,(N_g-N_g^0)
-e\,D_{Si}\,(N_{Si}-N_{Si}^0)\nonumber\\
&&\mbox{}+\frac{e^2\,N_c}{2 C_S}\,(N_S-N_S^0)^2\,
+\frac{e^2}{2 C_{Si}}\,(N_{Si}-N_{Si}^0)^2\,.
\label{eq01}
\end{eqnarray}
We have written the charge $Q_i$ of a given component $i$ as
$Q_i=-\,e(N_i-N_i^0)$, where $N_S$, $N_g$ and $N_{Si}$ are, respectively, the number of electrons in
a cluster (all clusters are equally charged), 
in graphene and in the Si layer in the equilibrium state in which these materials are in contact
and exchange charge, and $N_S^0$, $N_g^0$ and $N_{Si}^0$ are the same quantities in the uncharged 
state of these materials. Also, $D_S$, $D_g$ and $D_{Si}$ are the surface dipoles of
each substance. 

The conditions of thermodynamic equilibrium are obtained through the minimisation
of (\ref{eq01}) subjected to the constraints that the overall charge of the system 
is zero and that the potential difference between the two electrodes of the battery
is equal to $V_G$. The minimisation condition is thus given by
\begin{equation}
d{\cal E}=-eV_G\,dN_{Si}+\mu\,\left(N_c\,dN_S+dN_g+dN_{Si}\right)\,,
\label{eq02}
\end{equation}
since the charge transferred between the electrodes of the battery 
is $dq=-e\,dN_{Si}$, and where $\mu$ is the chemical potential of the 
system. One obtains from (\ref{eq02}) the following equilibrium conditions
\begin{eqnarray}
\overline{\mu}_g&=&\overline{\mu}_S+e(D_g-D_S)+\frac{e^2(N_S-N_S^0)}{C_S},
\label{eq03}\\
\overline{\mu}_g&=&\overline{\mu}_{Si}+e(D_g-D_{Si})+eV_G+\frac{e^2(N_{Si}-N_{Si}^0)}{C_{Si}},
\label{eq04}
\end{eqnarray}
where $\overline{\mu}_S=\frac{\partial G}{\partial N_j}\mid_{N_j=N_S}$,
with $N_j$ being the number of charges in cluster $j$,
$\overline{\mu}_g=\frac{\partial G}{\partial N_g}$,
$\overline{\mu}_{Si}=\frac{\partial G}{\partial N_{Si}}$,
are the chemical potentials of the different components of the system
in the absence of a dipole layer. These quantities are also
called the internal contribution of a metal to its work-function\cite{seitz}.
Note that $\mu=\overline{\mu}_g-eD_g$, i.e. the chemical potential of the
system is equal to the (full) chemical potential of the subsystem 
kept at zero voltage.

The two equilibrium conditions (\ref{eq03},\ref{eq04}) are not sufficient to determine
the level of doping of the constituents of the system. These equations have to be
supplemented with the neutrality condition for the overall system. 
One can write this condition as
\begin{equation}
N_c\,\Delta N_S+\Delta N_g+\Delta N_{Si}=0\,,
\label{eq05}
\end{equation}
with $\Delta N_S=N_S-N_S^0$, $\Delta N_g=N_g-N_g^0$ and $\Delta N_{Si}=N_{Si}-N_{Si}^0$.
The quantities that appear in (\ref{eq05}) are related to the variations of 
the carrier density of the cluster, of graphene, and of the Si
layer, by $\Delta N_S={\cal V}_S\,\delta n_S$, $\Delta N_g= A_g \,\delta n_g$
and $\Delta N_{Si}= A_g \,\delta n_{Si}$, where ${\cal V}_S$ is the volume of the cluster. 
The number of clusters is given by $N_c=N_{am}/n_{amc}$, where $n_{amc}$ is the
number of metal atoms per cluster, which is equal to $n_{amc}=z_S {\cal V}_S/v_S$,
with $v_S$ being the volume of the metallic unit cell and $z_S$ being the number 
of atoms in the unit cell ($z_S=1$ for Pt, $z_S=2$ for Ti). 
Expressing $N_{am}$ in terms of $c_S$, which was introduced at 
the beginning of this section, one obtains for $N_c$
\begin{equation}
N_c=\frac{c_S v_S A_g}{z_S {\cal V}_S A_c}\,,
\label{eq06}
\end{equation}
where we have expressed $n_u$ as the ratio between
the area of the graphene sheet and the unit cell area.
Substituting this formula in equation (\ref{eq05})
and expressing $\Delta N_S$, $\Delta N_g$ and $\Delta N_{Si}$
in terms of the variations of the charge density of each media,
one obtains
\begin{equation}
\frac{c_S v_S}{z_S A_c}\,\delta n_S+\delta n_g+\delta n_{Si}=0\,.
\label{eq07}
\end{equation}

The equations (\ref{eq03},\ref{eq04}), when written in terms of the
variations of density $\delta n_S$ and $\delta n_{Si}$, become
\begin{eqnarray}
\overline{\mu}_g&=&\overline{\mu}_S+e(D_g-D_S)+\frac{e^2 {\cal V}_S}{C_S}\,\delta n_S\,,
\label{eq012}\\
\overline{\mu}_g&=&\overline{\mu}_{Si}+e(D_g-D_{Si})+eV_G+\frac{e^2 b}{\epsilon}\,\delta n_{Si}\,,
\label{eq013}
\end{eqnarray}
These two equations, which impose the equality of the so-called electro-chemical
potentials\cite{madelung} between a metallic cluster and the graphene sheet, and between the
graphene sheet and the Si layer, when supplemented by (\ref{eq07}), are sufficient to 
determine the level of doping of graphene. However, we still
need to relate the chemical potential $\overline{\mu}_S$ to $\delta n_S$,
$\overline{\mu}_g$ to $\delta n_g$ and $\overline{\mu}_{Si}$ to $\delta n_{Si}$, in other
words, we need the equation of state for the different components of the system.
One writes $\Delta \varepsilon_F^{S}=\overline{\mu}_S-\varepsilon_F^{S}$ for
the Fermi energy variation of the cluster, $\Delta \varepsilon_F^{g}=\overline{\mu}_g-\varepsilon_F^{g}$
for the Fermi energy variation of graphene and $\Delta \varepsilon_F^{Si}=\overline{\mu}_{Si}-\varepsilon_F^{Si}$
for the Fermi energy variation of the Si layer, measured with respect to the uncharged 
ground state of each of these constituents. In the case of the clusters or of the 
Si layer, one has $\delta n_S\approx \rho_{S}(\varepsilon_F^{S})\,\Delta\varepsilon_F^{S}$ and
$\delta n_{Si}\approx \rho_{Si}(\varepsilon_F^{Si})\,\Delta\varepsilon_F^{Si}$
since the density of states $\rho(\varepsilon_F)$ is approximately constant
for these materials at the Fermi level. Substituting these definitions
in (\ref{eq012}) and (\ref{eq013}) and taking into account that the bulk work
functions of the transition metal, of graphene, and of Si, are given by
$W_{B}=eD_{B}-\varepsilon_F^{B}$, $W_g^0=eD_g^0-\varepsilon_F^g$ and 
$W_{Si}^0=eD_{Si}^0-\varepsilon_F^{Si}$\cite{lang1}, one obtains 
\begin{eqnarray}
\delta n_S\!\!&=&\!\!\frac{W_{B}\!-\! W_g^0-e(\Delta D_g\!-\!\Delta D_{S})\!-\!\zeta_S+\Delta\varepsilon_F^{g}}{
1/\rho_{S}(\varepsilon_F^{S})+e^2 {\cal V}_S/C_S},
\label{eq014}\\
\delta n_{Si}\!\!&=&\!\!\frac{W_{Si}^0\!-\!W_g^0+e(\Delta D_{Si}\!-\!\Delta D_g\!-\! V_G)+\Delta\varepsilon_F^{g}}{
1/\rho_{Si}(\varepsilon_F^{Si})+e^2 b/\epsilon},
\label{eq015}
\end{eqnarray}
where $\zeta_S=\varepsilon_F^{S}-\varepsilon_F^{B}$ is the difference 
between the Fermi energy of the TM cluster and the Fermi energy
of the bulk transition-metal and $\Delta D_S=D_S-D_{B}$,
$\Delta D_g=D_g-D_g^0$ and $\Delta D_{Si}=D_{Si}-D_{Si}^0$ are the
induced surface dipoles on each component of the system due to finite size
effects and to the presence of the other components. 
One can estimate $\rho_{Si}(\varepsilon_F^{Si})\approx
\frac{m^*_{Si}}{\pi\hbar^2}$, the result for a free two-dimensional
electron gas, where $m^*_{Si}\approx m_e$ is the electron's effective mass in Si.
With $b\approx 300$ nm, one has $\rho_{Si}(\varepsilon_F^{Si})\gg 
\epsilon/(e^2b)$ and one can neglect the first term in the denominator of 
(\ref{eq015}). One can thus write (\ref{eq014},\ref{eq015}) as
\begin{eqnarray}
\delta n_S&=&\frac{W_{B}-W_g^0-\Delta_c+\Delta\varepsilon_F^{g}}{
1/\rho_{S}(\varepsilon_F^{S})+e^2 {\cal V}_S/C_S}\,,
\label{eq016}\\
\delta n_{Si}&=&\frac{\epsilon}{e^2 b}\left[\,e\,(V_0-V_G)+\Delta\varepsilon_F^{g}
\,\right]\,,
\label{eq017}
\end{eqnarray}
where $\Delta_c=\zeta_S+e(\Delta D_g-\Delta D_{S})$ represents a
correction to the doping of the clusters due to their finite size
and to the induced surface dipoles, and $V_0=(W_{Si}^0-W_g^0)/e+\Delta
D_{Si}-\Delta D_g$.
The quantity $\Delta_c$ can be interpreted as giving the overall magnitude of
the {\it effective} chemical interaction between the clusters and the graphene sheet.

The density of states $\rho_{g}(\epsilon)$ of graphene is zero at the Dirac point and
one needs to consider its full functional form in that neighbourhood. It is approximately 
given by\cite{peresrmp}
\begin{eqnarray}
\rho_g(\varepsilon)&=&\frac{4}{\sqrt{3}\pi t^2 A_c}\,\mid \varepsilon-\varepsilon_F^g\mid\,,
\label{eq018}
\end{eqnarray}
where $t=2.7$ eV is the first-nearest neighbour hopping matrix element
in graphene. Note that the presence of impurities in graphene, either
intrinsic or the deposited TM atoms themselves, will modify
the density of states given in (\ref{eq018}) for high enough
impurity concentrations\cite{leconte}.
Integrating $\rho_g(\varepsilon)$ between the lower band limit
$\varepsilon_F^g-\sqrt{\sqrt{3}\pi}t$ and the Fermi energy $\varepsilon_F^g$ 
\cite{dora} yields
a result of two electrons per unitary cell.
 
Integrating $\rho_g(\varepsilon)$ between $\varepsilon_F^g$ and $\mu_g$, one
obtains for $\delta n_g$ the result
\begin{equation}
\delta n_{g}=
\pm\,\frac{2}{\sqrt{3}\pi t^2 A_c}\,(\Delta\varepsilon_F^{g})^2\,.
\label{eq019}
\end{equation}
with the plus sign if $\Delta\varepsilon_F^{g}>0$ and the minus sign otherwise.

Substituting equations (\ref{eq016}), (\ref{eq017}) and (\ref{eq019}) in 
(\ref{eq07}), we finally obtain a second-degree equation for $\Delta\varepsilon_F^{g}$
\begin{equation}
\pm\,(\Delta\varepsilon_F^{g})^2+\Lambda\,\Delta\varepsilon_F^{g}-\Omega=0\,,
\label{eq020}
\end{equation}
with the plus sign if $\Delta\varepsilon_F^{g}>0$ and negative sign otherwise, and where 
\begin{eqnarray}
\Lambda&=&\frac{\sqrt{3}\pi t^2}{2}\left[\frac{c_S
    v_S}{z_S(1/\rho_{S}(\varepsilon_F^{S})
+e^2 {\cal V}_S/C_S)}+\frac{\epsilon A_c}{e^2 b}\right],
\label{eq021}\\
\Omega&=&\frac{\sqrt{3}\pi
  t^2}{2}\left[\frac{c_Sv_S\,(W_g^0+\Delta_c-W_{B}\,)}{
z_S(1/\rho_{S}(\varepsilon_F^{S})+e^2 {\cal V}_S/C_S)}\right.\nonumber\\
&&\left.+\frac{\epsilon A_c}{e\,b}\,(\,V_G-V_0\,)\right]\,.
\label{eq022}
\end{eqnarray}

If we take the positive sign in equation (\ref{eq020}) then a positive
solution exists if $\Omega>0$. Conversely, if we take the negative sign 
in this equation, a negative solution exists if $\Omega<0$. 
One can thus write for $\Delta\varepsilon_F^{g}$, the solution \cite{noteccl}
\begin{equation}
\Delta\varepsilon_F^{g}=\mbox{sign}(\Omega)\,\left(\,\sqrt{\frac{\Lambda^2}{4}+\mid \Omega\mid}-\frac{\Lambda}{2}\,\right)\,.
\label{eq023}
\end{equation}

One can see from equation (\ref{eq023}) that for a given concentration $c_S$, 
one can, through the application of a gate-voltage $V_D$ such that $\Omega=0$, bring the 
graphene sheet to its uncharged state, as $\Delta\varepsilon_F^{g}=0$. The gate voltage 
$V_D$ can be determined from transport measurements on Pt or Ti-covered
graphene\cite{pi}, since the conductivity will display the minimum
characteristic of the Dirac point for that applied voltage. Likewise, the gate 
voltage $V_0$ can be determined from the same measurements performed on
the uncovered graphene samples, since it follows from equation
(\ref{eq023}) that for $c_S=0$, $\Delta\varepsilon_F^{g}=0$ at $V_G=V_0$. Thus,
the conductivity will also display the characteristic minimum at this applied
voltage. Therefore, one can extract $\Delta_c$ from experiment. It is given by
\begin{eqnarray}
\Delta_c&\!=\!&W_{B}-W_g^0\nonumber\\
&&\mbox{}-\frac{\epsilon z_S A_c
(1/\rho_{S}(\varepsilon_F^{S})+e^2{\cal V}_S/C_S)(V_D-V_0)}{e\,b\, c_S v_S}.
\label{eq025}
\end{eqnarray}

The density of states at the Fermi level of bulk Pt or bulk Ti can be extracted from specific heat
measurements\cite{andresen,estermann}, through $\rho(\varepsilon_F^{B})=3\gamma/(\pi k_B)^2$, using
the general result from Landau's Fermi liquid theory, where $\gamma$ is the linear coefficient
for the dependence of the electronic specific heat $c_V$ on the temperature.
Finite size corrections to the bulk density of states may be estimated
for a spherical cluster, using the results for a free-electron gas
\cite{hill}, as $\rho_{S}(\varepsilon_F^{S})=\rho(\varepsilon_F^{B})-3m^*/(8\pi \hbar^2 R)$,
where $m^*_{Pt}\approx 2m_e$\cite{ketterson}, $m^*_{Ti}\approx
3.15m_e$\cite{estermann} is the electron's effective mass in platinum
or titanium and $R$ is the radius of the cluster. Also, for a spherical cluster, 
${\cal V}_S=4\pi R^3/3$ and $C_S$ can be written as a power series on a
parameter dependent on the cluster radius and on the distance $L$ of its
centre to the graphene sheet (see appendix \ref{secA}). 
We take $W_g^0=4.5$ eV as in \cite{pi}, $W_{Pt}^0=5.64$ eV and
$W_{Ti}^0=4.33$ eV for polycrystalline platinum and polycrystalline titanium,
and $W_{Si}^0=4.6$ eV\cite{CRC}. We note that the model as defined contains
two unknown parameters, namely $R$ and $L$.
\begin{table}[t]
\begin{center}
\begin{tabular}{|c|c|c|c|c|c|} 
\hline $c_S$ & $V_D$ (V) & $R$ (nm)&$\Delta_c$ (eV)& $\Delta_c^{est}$ (eV)&
 $p_S$\\ \hline
 0.025&-11.4& 0.6&2.47&2.55&-0.014\\ \hline
 0.071& -29.0& 0.6&2.49&2.55&-0.014\\ \hline 
 0.127& -46.0&  0.6&2.37&2.55&-0.013\\ \hline
\end{tabular}
\end{center}
\caption{Values of the concentration of Pt atoms per unit cell of graphene $c_S$
and applied voltages $V_D$ corresponding to the minimum of conductivity 
for the Pt-1 sample studied in \cite{pi} ($V_0=-1.94$ V). The values of $\Delta_c$ were determined
using formula (\ref{eq025}). We have also computed the number of electrons
$p_S$ per Pt atom at the Dirac point. This number was estimated in \cite{pi}
as -0.014 e/Pt atom.}
\label{t1}
\end{table}

\begin{table}[t]
\begin{center}
\begin{tabular}{|c|c|c|c|c|c|} 
\hline $c_S$ & $V_D$ (V) & $R$ (nm)&$\Delta_c$ (eV)& $\Delta_c^{est}$ (eV)&
 $p_S$\\ \hline
 0.0065&1.56& 0.6&2.16&2.55&-0.011\\ \hline
 0.019& -2.56& 0.6&2.26&2.55&-0.012\\ \hline 
 0.039& -11.0&  0.6&2.45&2.55&-0.014\\ \hline
0.064& -24.3&  0.6&2.66&2.55&-0.016\\ \hline
\end{tabular}
\end{center}
\caption{Values of the concentration of Pt atoms per unit cell of graphene $c_S$
and applied voltages $V_D$ corresponding to the minimum of conductivity 
for the Pt-3 sample studied in \cite{pi} ($V_0=3.41$ V). The values of $\Delta_c$ were determined
using formula (\ref{eq025}). We have also computed the number of electrons
$p_S$ per Pt atom at the Dirac point. This number was estimated in \cite{pi}
as -0.019 e/Pt atom.}
\label{t1A}
\end{table}

\begin{table}[t]
\begin{center}
\begin{tabular}{|c|c|c|c|c|c|} 
\hline $c_S$ & $V_D$ (V) & $R$ (nm)&$\Delta_c$ (eV)& $\Delta_c^{est}$ (eV)&
 $p_S$\\ \hline
 0.0038&-18.6&0.19&1.70&1.83&-0.179\\ \hline
 0.0077&-41.0&0.19&1.89&1.83&-0.198\\ \hline 
 0.0115&-61.0&0.19&1.89&1.83&-0.198\\ \hline
 0.0153&-73.5&0.19&1.71&1.83&-0.180\\ \hline
 0.0191&-82.4&0.19&1.52&1.83&-0.161\\ \hline
\end{tabular}
\end{center}
\caption{Values of the concentration of Ti atoms per unit cell of graphene $c_S$
and applied voltages $V_D$ corresponding to the minimum of conductivity 
for the Tt-1 sample studied in \cite{pi} ($V_0=-0.57$ V). The values of $\Delta_c$ were determined
using formula (\ref{eq025}). We have also computed the number of electrons
$p_S$ per Ti atom at the Dirac point. This number was estimated in \cite{pi}
as -0.174 e/Ti atom.}
\label{t2}
\end{table}
Using these results, as well as the values
of $c_S$, $V_0$ and $V_D$ measured by \cite{pi} 
(raw data is a courtesy of Kawakami's group) and taking the radius of the cluster
to be $0.6$ nm for Pt\cite{radiusPt} and $0.188$ nm for Ti
\cite{radiusTi} and the distance from the centre of the cluster to the plane
equal to $0.85$ nm for Pt and $0.44$ nm for Ti, one obtains 
for $\Delta_c$ the results given in tables \ref{t1}, \ref{t1A} and \ref{t2}. 
Note that one does not dispose of direct information (e.g. from AFM
measurements) regarding the values of $R$ and $L$.
The values indicated above were chosen such as to provide agreement
between the values of $\Delta_c$ and $\Delta_c^{est}$ in tables \ref{t1},
\ref{t1A} and \ref{t2}, as well as  between the theoretical and experimental
asymptotic values for the contribution to the resistivity coming from the
presence of the clusters\cite{LmR} (see section \ref{secB}). We have
considered here and below the measurements made with samples Pt-1, 
Pt-3 and Ti-1 (in the notation of \cite{pi}), since these samples, when
uncovered, presented the smallest values of $V_0$ measured, indicating a low 
level of intrinsic disorder.

One can estimate the correction to the Fermi energy, due to the cluster's finite radius,
from the free-electron gas result, as $\zeta_S\approx\frac{3\pi^2\hbar^4}{8 (m^{*})^2 R}\,
\cdot\rho(\varepsilon_F^{B})$.
Using the result quoted in the references \cite{gio1,gio2} for the induced dipole
$e\,(\Delta D_g-\Delta D_S)\approx 0.9$ eV\cite{NoteFS}, one obtains for $\Delta_c^{est}$
the results presented in tables \ref{t1}, \ref{t1A} and \ref{t2} for Pt and Ti,
respectively. Thus, it is seen that the larger
value of the chemical interaction $\Delta_c$ with respect to the case studied
in\cite{gio1,gio2} (particularly in platinum that has a larger DOS at the
Fermi level) is due to a large shift of the Fermi energy of the clusters
with respect to that of the bulk TM metal, caused by their finite radius.

\begin{figure}
\begin{center} 
\includegraphics[clip,width=7cm]{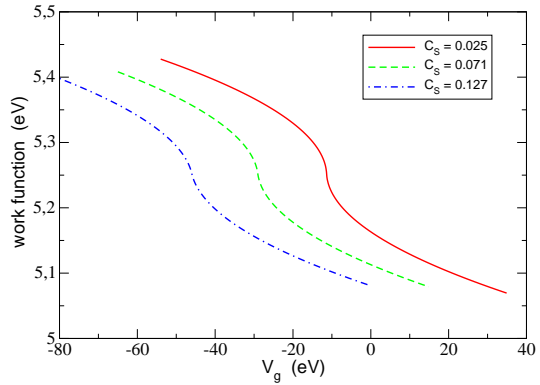}
\par
\end{center}
\caption{Ionisation potential of cluster as a function of the applied gate voltage,
for the Pt-1 sample, with coverages $c_S=0.025, 0.071$ and $0.127$ ML (rgb).}
\label{fig:3} 
\end{figure}

\begin{figure}
\begin{center} 
\includegraphics[clip,width=7cm]{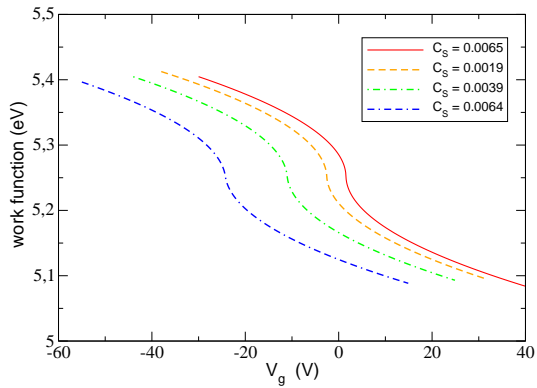}
\par
\end{center}
\caption{Ionisation potential of cluster as a function of the applied gate voltage,
for the Pt-3 sample, with coverages $c_S=0.0065,\,0.019,\,0.039$ and $0.064$ ML (rogb).}
\label{fig:3Pt3} 
\end{figure}

\begin{figure}
\begin{center} 
\includegraphics[clip,width=7cm]{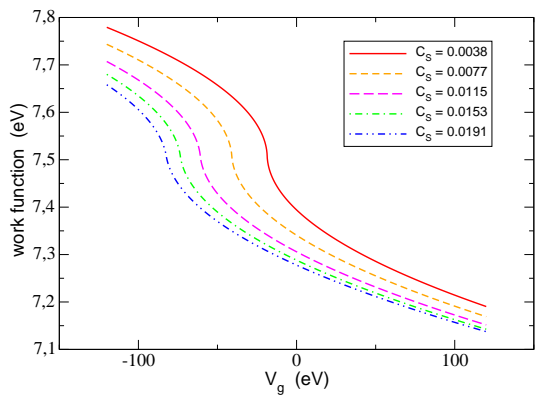}
\par
\end{center}
\caption{Ionisation potential of cluster as a function of the applied gate voltage,
for the Ti-1 sample, with coverages $c_S=0.0038,\, 0.0077,\, 0.0115,\, 0.0153$ 
and $0.0191$ ML (romgb).}
\label{fig:3Ti} 
\end{figure}

For a small cluster, the concept of work-function is ill-defined\cite{hyper},
as this quantity depends on the cluster's charge. One then speaks,
respectively, of the cluster's ionisation potential if one is withdrawing an electron
from a cluster at equilibrium, or of the cluster's electron affinity, if the
electron is withdrawn from a negatively over-charged cluster. 
In appendix \ref{appO}, we compute the ionisation potential of a 
cluster based on a thermodynamic argument. We obtain 
from (\ref{eqapA4}) the result 
\begin{equation}
I_S=W_g^0+e\Delta D_g+\frac{e^2}{2C_S}-\Delta\varepsilon_F^{g}\,,
\label{eq024}
\end{equation} 
for the ionisation potential of the metallic cluster, where $\Delta\varepsilon_F^{g}$
is given by (\ref{eq023}). Substituting in (\ref{eq024}) the parameters 
as computed in table \ref{t1}, we have plotted in Figures \ref{fig:3}, 
\ref{fig:3Pt3} and \ref{fig:3Ti}
the result (\ref{eq024}) as function of the applied voltage, 
for the different coverages $c_S$ considered in \cite{pi}, for their Pt-1, Pt-3 and
Ti-1 samples\cite{ioeTi}. In these plots, we have ignored the (unknown) constant $e\Delta
D_g$\cite{objgio}. Nevertheless, such constant shift should be obtainable from a plot
of the experimental ionisation potential, and so provide an estimate of
$e\Delta D_g$.

It is also shown in appendix \ref{appO} that the electron affinity $A_S$ of a
metallic cluster is given by $A_S=I_S-e^2/C_S$, with
$W_g=\frac{1}{2}(I_S+A_S)$. Thus, $A_S<W_g<I_S$. Since the transfer of an
electron from the cluster to graphene would cost an energy $I_S-W_g>0$
and, conversely, the transfer of an electron from graphene to the cluster
would cost the same energy $W_g-A_S>0$, one sees that the equilibrium
state defined by equations (\ref{eq07}) to (\ref{eq013}) is indeed a stable
one. The theory exposed in this section constitutes the main result of this
paper.
\section{Scattering of electrons by the metallic clusters and 
its contribution to the resistivity in the FBA}
\label{secB}
In the previous section, we have computed the level of doping of a graphene 
sheet due to the presence of metallic clusters. We have also computed the
ionisation potential of a single cluster. These properties are equilibrium properties. 
However, the experiments of \cite{pi} measured the dependence of the
conductivity of graphene on the doping induced by the metallic clusters
and by the applied gate voltage.
In order to describe such dependence, one needs to determine 
the scattering potential on individual carriers due to the presence of the clusters. 
We will determine such a potential for spherical clusters, in an electrostatic approximation,
which allows for the use of the method of images.
Note that in such approximation, the graphene sheet is an equipotential surface. Therefore, the
Coulomb interaction due to the surface charge distribution of the clusters is perfectly screened
by the surface charge distribution that it is generated on the graphene sheet. However, such induced
charge distribution is spatially varying and it will thus correspond to a local variation of the
Fermi level of graphene \cite{fogler}. Such variation will enter in the Dirac
equation that describes the low-energy properties of graphene as a scattering
potential and will give rise to a variation of the conductivity. Note that the
local variation of the Fermi energy due to a cluster is not a parameter of the
model as in \cite{kats}, but depends on the level of doping of graphene.

The clusters, of radius $R$, are placed at a distance $L>R$ above the graphene sheet.
Each cluster possesses a charge $Q_S=-e\Delta N_S=-\frac{4}{3}\pi R^3 e\,\delta n_S$. 
We place the origin of the coordinate axis aligned with the centre of the
spheres, such that the graphene sheet is located at $z=-L$. 
In the subspace $z\geq -L$, the electrostatic potential can be approximately described
by the superposition 
\begin{equation}
V(\vecg{r})=\sum_i\,v(\vecg{r}-\vecg{r}_i)\,,
\label{eqB1}
\end{equation} 
of the potentials due to the individual clusters, located at $ \vecg{r}_i=(x_i,y_i,0)$.

In appendix \ref{secA}, we will show how $v(\vecg{r})$ can be written in terms of a series of
image charges, located in the cluster, and their images, located below the graphene plane. 
These charges depend on $Q_S$, $R$ and $L$ through a recursion relation.The
displacement field in the subspace $z\geq -L$ is given by $\vecg{D}=-\epsilon_0\,\nabla V(\vecg{r})$, whereas it is
equal to $\vecg{D}=-e\,\delta n_{Si}\,\vecg{e}_z$ in the space between the graphene sheet and 
the Si layer. The discontinuity of its normal component at $z=-L$ determines
the local density of charge $\sigma_g(x,y)$ in the graphene sheet. In terms of
the density of carriers $\delta n_g(x,y)=-\sigma_g(x,y)/e$, one has 
\begin{equation}
\label{eqB3}
\delta n_g(x,y)=-\,\delta n_{Si}+\frac{\epsilon_0}{e}\,\left.\frac{\partial
    V(\vecg{r})}{\partial z}\right|_{-L}\,,
\end{equation}
where the derivative with respect to $z$ is evaluated at the location
of the graphene plane, $z=-L$.
The spatial average of $\delta n_g(x,y)$ is given by equation (\ref{eq07}). 
Averaging equation
(\ref{eqB3}), we thus obtain $\left.\overline{\frac{\partial V(\vecb{r})}{\partial z}}\right|_{-L}=-\frac{e c_S v_S}{\epsilon_0 z_S A_c}
\,\delta n_S$. Furthermore, one can write equation (\ref{eqB3}) as
\begin{equation}
\label{eqB4}
\delta n_g(x,y)=\delta n_g+\frac{\epsilon_0}{e}\,\left[\left.\frac{\partial V(\vecg{r})}{\partial z}\right|_{-L}
-\left.\overline{\frac{\partial V(\vecg{r})}{\partial z}}\right|_{-L}\right]
\,.
\end{equation}
The precise form of the local density of carriers depends on the location of
the metallic clusters. However, assuming that one can treat clusters as independent
entities, one has that in the neighbourhood of a given cluster, located
at the origin of the coordinates, one can approximate (\ref{eqB4}) by
\begin{equation}
\label{eqB5}
\delta n_g(x,y)\approx \delta n_g+\frac{\epsilon_0}{e}\,\left[\left.\frac{\partial v(\vecg{r})}{\partial z}\right|_{-L}
-\left.\overline{\frac{\partial v(\vecg{r})}{\partial z}}\right|_{-L}\right]
\,,
\end{equation}
where $v(\vecg{r})$ is given by (\ref{eqB2}). Using the previous results, one can also estimate that 
\begin{equation}
\left.\overline{\frac{\partial v(\vecg{r})}{\partial z}}\right|_{-L}=-\frac{e c_S v_S}{\epsilon_0 z_S A_c N_c}
\,\delta n_S\,,
\label{eqB5a}
\end{equation}
where $N_c$ is the number of clusters. Since such a number is supposed to be very large, this
term is negligible and one has that
\begin{equation}
\label{eqB6}
\delta n_g(x,y)\approx \delta n_g+\frac{\epsilon_0}{e}\,\left.\frac{\partial v(\vecg{r})}{\partial z}\right|_{-L}
\,.
\end{equation}
Substituting equations (\ref{eqB2}) and (\ref{eqB2A}) 
for $v(\vecg{r})$ in (\ref{eqB6}) and expressing $Q_S$ in terms of $\delta n_S$ 
as above, one obtains the following series for $\delta n_g(r)$\cite{Note4}
\begin{eqnarray}
\label{eqB8}
\delta n_g(r)&=&\delta n_g-\frac{2\,R^3\,\delta
  n_S\,\sqrt{L^2-R^2}}{3\,g(\lambda,1)}\,\sum_{n=1}^\infty\,\frac{\lambda^n(1+\lambda^{2n})}{(1-\lambda^{2n})^2}
\nonumber\\
&&\mbox{}\times\frac{1}{\left[\,r^2\,+(L^2-R^2)\,\left(\frac{1+\lambda^{2n}}{1-\lambda^{2n}}\right)^2\,\right]^{3/2}}\,.
\end{eqnarray}
where $\lambda=\frac{1}{R}\,(L-\sqrt{L^2-R^2})$ and where $g(\lambda,1)$
is given by (\ref{eq6A}).

\begin{figure}
\begin{center} 
\includegraphics[clip,width=7cm]{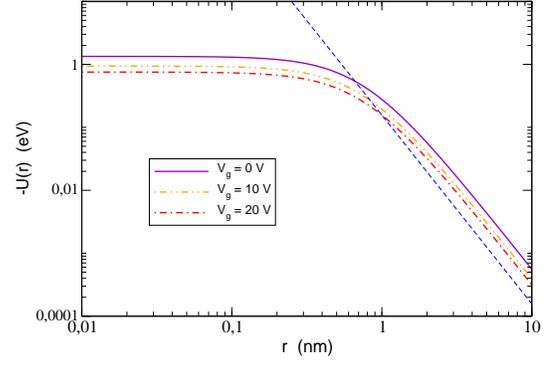}
\par
\end{center}
\caption{Log-log plot of scattering potential due to clusters for Pt-1 sample with coverage 
$c_S=0.025$ and applied voltages $V_G=0, 10, 20$ V. The dashed line
has slope $-3$ and is a guide to the eye.}
\label{fig:4} 
\end{figure}

The local variation of the carrier density $\delta n_g(r)$ in graphene
is related to the local variation of the Fermi energy $\Delta \varepsilon_F^g(r)$
through equation (\ref{eq019}). Assuming that one is far from the neutrality
point of graphene and that $\Delta \varepsilon_F^g(r)$ and $\Delta \varepsilon_F^g$
have the same sign, one has that the 
difference $U(r)=\Delta \varepsilon_F^g-\Delta \varepsilon_F^g(r)$
is given approximately by
\begin{eqnarray}
\label{eqB9}
U(r)&\approx&
\frac{\pi t^2\,A_c\, \delta
  n_S\,R^3\,\sqrt{L^2-R^2}}{2\sqrt{3}\,g(\lambda,1)\mid \Delta
  \varepsilon_F^g\mid}\,
\sum_{n=1}^\infty\,\frac{\lambda^n(1+\lambda^{2n})}{(1-\lambda^{2n})^2}
\nonumber\\
&&\mbox{}\times\frac{1}{\left[\,r^2\,+(L^2-R^2)\,\left(\frac{1+\lambda^{2n}}{1-\lambda^{2n}}\right)^2\,\right]^{3/2}}\,,
\end{eqnarray}
where $\Delta \varepsilon_F^g$ is given by 
(\ref{eq023}) and $\delta n_S$ is given by (\ref{eq016}). 
The function $U(r)$ is the electron
scattering potential due to a single cluster and depends,
in this approximation, on the cluster carrier density $\delta n_S$
and also on the level of doping of graphene itself (through
its dependence on $\mid\Delta\varepsilon_F^g\mid$). Note
that $U(r)$ is attractive if $\delta n_S<0$ (the cluster is doped
with holes, as seen in experiment), as one would expect. The plot
of $U(r)$ is given, for different values of the gate voltage $V_G$, in
figure \ref{fig:4}.

We can also compute from (\ref{eqB8}), for later use, the size of the region
in graphene that contains a charge $-Q_S$ of equal magnitude to that of an
individual clusters, in the case in which graphene is electron-doped (i.e.
for $V>V_D$). We integrate equation (\ref{eqB8}) within the disk $r<R_S$, 
the region whose size we wish to calculate. We obtain after some
cancellations, the following equation for $R_S$
\begin{eqnarray}
\label{eqB9A}
R_S^2&=&\frac{4 R^3\sqrt{L^2-R^2}\mid \delta n_S\mid}{3 g(\lambda,1)\,\delta n_g}\,
\sum_{n=1}^\infty\,\frac{\lambda^n(1+\lambda^{2n})}{(1-\lambda^{2n})^2}\nonumber\\
&&\mbox{}\times
\frac{1}{\left[\,R_S^2\,+(L^2-R^2)\,\left(\frac{1+\lambda^{2n}}{1-\lambda^{2n}}\right)^2\,\right]^{1/2}}\,.
\end{eqnarray}
This quantity should be compared with the average distance $R_c$ between clusters,
that can be simply defined through the relation $\pi R_c^2=A_g/N_c=\frac{4\pi
  R^3 A_c z_S}{3 c_S v_S}$, i.e. in
terms of the average area per cluster. If one defines the ratio $\xi=R_S/R_c$
between these two quantities, one has, noting that the second fraction in
the infinite sum of (\ref{eqB9A}) can be simply approximated by $R_S^{-1}$ if
$R_S\gg L$, that $\xi$ is approximately given by
\begin{eqnarray}
\label{eqB9B}
\xi&=&\left(\,\sqrt{\frac{3}{4}} \frac{v_S^{3/2}\,\sqrt{L^2-R^2}}{(A_c z_S R)^{3/2}\, g(\lambda,1)}\,
\sum_{n=1}^\infty\,\frac{\lambda^n(1+\lambda^{2n})}{(1-\lambda^{2n})^2}
\,\right)^{1/3}\nonumber\\
&&\mbox{}\times\,\left(\frac{c_S^{3/2}\mid \delta
      n_S\mid}{\delta n_g}\right)^{1/3}\,,
\end{eqnarray}
where the first term is a constant for fixed $L$ and $R$.
The validity of the independent cluster approximation, assumed above when
passing from (\ref{eqB4}) to (\ref{eqB5}), depends on the condition 
$\xi<1$\cite{Fulde} being fulfilled.

The contribution of the metallic clusters to the resistivity of the sample
of graphene can be easily computed within the FBA, once the scattering potential
is known. Within the semi-classical theory\cite{tapash} based on the Boltzmann
equation, the contribution of the clusters to the conductivity of the sample
is given by
\begin{equation}
\label{eqB10}
\sigma_{cl}=\frac{e^2 v_F k_F \tau_{cl}(k_F)}{\pi \hbar}\,
\end{equation} 
where $v_F=\frac{\sqrt{\sqrt{3} A_c/2}\,t}{\hbar}$ is the Fermi velocity in graphene,
$k_F=\frac{\mid\Delta\varepsilon_F^g\mid}{\hbar v_F}$ is the momentum of a quasi-particle
at the Fermi surface of doped graphene, measured with respect to the Dirac point,
and $\tau_{cl}(k_F)$ is the transport lifetime of a quasi-particle at the Fermi surface, due
to the scattering with the metallic clusters. The expression above already accounts for
the double spin and valley degeneracy (existence of two independent Dirac points).
Equation (\ref{eqB10}) is known to apply as long as $k_F l_e\gg 1$, where $l_e$ is
the electron mean-free path. This condition holds in the diffusive regime where graphene 
is highly-doped (i.e. far away from the Dirac point) and for low-impurity concentration.
The inverse of $\tau_{cl}(k_F)$ can be computed 
in the FBA, by the application of Fermi's Golden-Rule
\begin{eqnarray}
\label{eqB11}
\frac{1}{\tau_{cl}(k_F)}&=&\frac{2\pi}{\hbar}\,N_c\,\sum_{\vecg{k}'}\mid\bra{\vecg{k}'}\,U(r)\,\ket{\vecg{k}}\mid^2
\nonumber\\
&&\mbox{}\times (1-\hat{\vecg{k}}'\cdot\hat{\vecg{k}})\,\delta(\varepsilon_{\vecb{k}'}-\varepsilon_{\vecb{k}})\,,
\end{eqnarray}
where $N_c$ is as above the number of clusters, i.e. the number of scattering centres, $\mid\!
\vecg{k}\!\mid\,=k_F$, $\varepsilon_{\vecb{k}}=\hbar v_F\mid\!
\vecg{k}\!\mid$, $\varepsilon_{\vecb{k}'}
=\hbar v_F\mid\! \vecg{k}'\!\mid$, $\hat{\vecg{k}}$,
$\hat{\vecg{k}}'$ are the unit vectors in the direction of $\vecg{k}$ and $\vecg{k}'$
and $U(r)$ is given by \eqref{eqB9}.

In \eqref{eqB11}, one needs to take into account the spinorial nature of the wave-functions $\ket{\vecg{k}}$,
$\ket{\vecg{k}'}$. The spinor $u_{\vecb{k}}(\vecg{r})=\langle \,\vecg{r}\,\ket{\vecg{k}}$,
which is normalized over the area $A_g$ of the sample is given by
\begin{equation}
\label{eqB12}
u_{\vecb{k}}(\vecg{r})=\frac{1}{\sqrt{2A_g}}\,
\left(\begin{array}{c}
e^{\mbox{$-i\theta_{\vecb{k}}/2$}}\\
\pm e^{\mbox{$i\theta_{\vecb{k}}/2$}}
\end{array}\right)
\,e^{i\vecb{k}\cdot\vecb{r}}\,,
\end{equation}
where $\tan \theta_{\vecb{k}}=k_y/k_x$. The $\pm$ signs stand for states
with the same momentum and opposite energies relative to the Dirac point.
The expression for $u_{\vecb{k}'}(\vecg{r})$ is entirely analogous.

\begin{figure}
\begin{center} 
\includegraphics[clip,width=7cm]{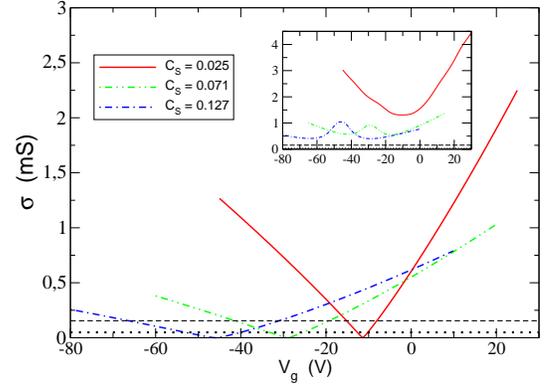}
\par
\end{center}
\caption{Contribution to the conductivity of graphene sample due to
  Pt-clusters (Pt-1 sample) as function of the applied gate voltage, 
for the coverages $c_S=0.025, 0.071$ and $0.127$ ML (rgb), following the
prediction of equation (\ref{eqB14}). The values
of the two fitting parameters, the cluster radius $R=0.6$ nm and distance 
$L=0.85$ nm between the  cluster center and the graphene sheet, are as given 
in Table \ref{t1}. Also shown are the minimal value for the conductivity
$4e^2/(\pi h)$ predicted by the SCBA and the value of $4e^2/h$ measured by
\cite{geim07,tan07}. Inset: contribution to the conductivity from the clusters
as measured by \cite{pi}.}
\label{fig:5} 
\end{figure}

Substituting (\ref{eqB12}) and the analogous expression for $u_{\vecb{k}'}(\vecg{r})$
in \eqref{eqB11}, converting the summation over
$\vecg{k}'$ into an integral, performing  the integral over $k'$ 
using the delta function and expressing $N_c/A_g=\frac{3 c_S v_S}{4\pi R^3 z_S A_c}$,
we obtain the following result in terms
of an angular integral over the scattering angle $\phi=\theta_{\vecb{k}'}-\theta_{\vecb{k}}$,
\begin{eqnarray}
\label{eqB13}
\frac{1}{\tau_{cl}(k_F)}&\!=\!&\frac{\pi^2c_S v_S v_F (\delta n_S)^2 R^3}{12 z_S A_c k_F g^2(\lambda,1)}
\int_0^{2\pi}\,d\phi\sin^2 \phi\\
&&\mbox{}\times\left[\,\sum_{n=1}^\infty\frac{\lambda^n}{1-\lambda^{2n}}
e^{-\frac{2\,k_F\,\sqrt{L^2-R^2}\,(1+\lambda^{2n})\,\sin(\phi/2)}{1-\lambda^{2n}}}
\right]^2.\nonumber
\end{eqnarray}
The expression \eqref{eqB13}, as it stands, 
cannot be written in terms of elementary functions. However,
the resulting integral is elementary if one can substitute 
the exponential functions in the infinite sum by $1$, i.e. if their exponents
are very small. The largest exponent is the one coming from the term with
$n=1$ and is equal to $2k_F L \sin(\phi/2)$. Thus, this approximation
is valid if $k_F L\ll 1$. Since $\delta n_g=\pm \frac{k_F^2}{\pi}$,
one can write the above condition as $\mid \!\delta n_g\!\mid\ll \frac{1}{\pi L^2}$.
Taking $L\approx 1$ nm, one obtains from this condition that $\mid \!\delta n_g\!\mid\ll 
10^{13}-10^{14} \,\mbox{e}/\mbox{cm}^2$. This condition is obeyed for all experimentally 
applied voltages. Thus, substituting the resulting expression for $\tau_{cl}(k_F)$
in \eqref{eqB10}, one obtains 
\begin{equation}
\label{eqB14}
\sigma_{cl}(k_F)=\sigma_{min}\,\frac{6 \,z_S\,A_c\,\mid\!\delta n_g\!\mid}{
\pi \,c_S\,v_S\,R^3\,(\delta n_S)^2}\,,
\end{equation}
where $\sigma_{min}=4e^2/(\pi h)$ is the minimal conductivity of undoped graphene.
Using the equations \eqref{eq023}, \eqref{eq019} and \eqref{eq016} for
$\Delta \varepsilon_F^g$, $\delta n_g$ and $\delta n_S$ in \eqref{eqB14},
with the relevant parameters in these equations taking the values 
as given in tables \ref{t1}, we plot below (see figure \ref{fig:5})
the contribution to the conductivity of the graphene sample due to 
the clusters for the different coverages considered in \cite{pi} 
for their Pt-1 sample, as a function of the applied gate voltage $V_G$
and as predicted by equation (\ref{eqB14}).
Note the slight asymmetry of the curves with respect to the Dirac point,
due to the variation of the carrier density $\delta n_S$ of
the cluster with the applied gate voltage.

One can also easily compute from (\ref{eqB14}) the contribution of the
clusters to the mobility of the samples, defined as
$\mu_{cl}=\sigma_{cl}/(e\mid\delta n_g\mid)$. One obtains 
\begin{equation}
\label{eqB14A}
\mu_{cl}^{-1}=\frac{h\pi c_S}{32e A_c}n_{amc}p_S^2\,,
\end{equation}
where $p_S=v_S\delta n_S/z_S$ is the number of electrons per TM atom or doping
efficiency. 

The comparison of the result obtained in \eqref{eqB14} with the experimental
results of Pi et al. requires that we extract from their experimental data for
the overall resistivity of the sample, the contribution coming solely from the 
metallic clusters, since there are other types of scatterers contributing
to the resistivity even in an uncovered sample, as discussed in the introduction.
Therefore, we make the following hypothesis regarding the dependence of 
the overall resistivity of a sample of graphene on the applied
gate voltage $V_G$ and on the metallic coverage $c_S$
\begin{eqnarray}
\rho(V_G,c_S,n_i)&=&\rho_{imp}(\delta
n_g,n_i)+c_S\,\tilde{\rho}_{cl}(\delta n_g)\nonumber\\
&&\mbox{}+\rho_{MS}(\delta n_g,c_S,n_i)\,,
\label{eqB15}
\end{eqnarray}
where $n_i$ is the concentration of intrinsic impurities $n_i$ in the sample.
The function $\rho_{imp}(\delta n_g, n_i)$ is the contribution to the resistivity coming
from the intrinsic impurities of the sample, which we take to be a sole function
of the doping and of $n_i$. This function can be extracted from the measurements
done at zero coverage by expressing $\rho(V_G,0,n_i)$ as a function of $\delta n_g$,
using equations \eqref{eq019} and \eqref{eq023} with $c_S=0$.
The second term is the contribution to the conductivity due to 
scattering by a single cluster and is therefore linear in $c_S$.
Equation \eqref{eqB14} is of this form, as we can always express
$\delta n_S$ in it in terms of $\delta n_g$ through \eqref{eq016}
and \eqref{eq019}. However, this equation predicts an infinite resistivity 
due to the clusters at the Dirac point, since it assumes the system
to be in the diffusive regime, an assumption that fails close to
the Dirac point, as discussed above (see also \cite{leconte}, where
a similar situation occurs). Since the samples show a 
finite conductivity at the Dirac point and at finite coverage,
we cannot take \eqref{eqB14} as it stands. Instead, we write for
$\tilde{\sigma}_{cl}=\tilde{\rho}_{cl}^{-1}$, the ansatz 
\begin{equation}
\label{eqB16}
\tilde{\sigma}_{cl}=\tilde{\sigma}^0_{cl}+\sigma_{min}\,\frac{6 \,z_S\,A_c\,\mid\!\delta n_g\!\mid}{
\pi \,v_S\,R^3\,(\delta n_S)^2}\,,
\end{equation}
where $\tilde{\sigma}^0_{cl}$ is an extra contribution to the 
conductivity due to the clusters, which acts as an additional fitting 
parameter. Finally, $\rho_{MS}(\delta n_g,c_S,n_i)$ is 
the contribution to the resistivity due to multiple scattering events 
involving the metallic clusters, be it multiple scattering by a single
cluster, scattering events involving different clusters, or events involving 
clusters and intrinsic impurities in graphene, and is a general function of 
$c_S$, $\delta n_g$ and $n_i$. 

If one were to assume that $\rho_{MS}(\delta n_g,c_S,n_i)$ were
negligible, the function $(\,\rho(V_G,c_S,n_i)-\rho(V_G,0,n_i)\,)/c_S$,
expressed in terms of $\delta n_g$, would be independent of  
the coverage $c_S$, i.e. it would be a universal curve. This is not the case 
for the metallic coverages considered by Pi et al, since these coverages are simply
too large for multiple-scattering to be neglected, as will be shown below. 
In figure \ref{fig:6}, we perform a fitting of the theory to the results 
obtained with sample Pt-1 at the lowest coverage studied, but the objective of
such fitting is merely to show that with the parameters characterising the
clusters as given in table \ref{t1}, the theoretical and experimental results
have the same order of magnitude and show the same asymptotic behaviour.
The purple continuous plot represents the inverse of the 
function given by \eqref{eqB16}, with $\tilde{\sigma}^0_{cl}$ chosen so that
the maximum of this curve and the maximum of the red-dashed curve coincide (this is the
only free fitting parameter). We see that one is able to reproduce the asymptotic behaviour
of the experimental curve at large doping. The same asymptotic behaviour is also
observed at higher coverages, but the curves deviate significantly from the
universal curve hypothesis in the neighbourhood of the Dirac point.
\begin{figure}
\begin{center} 
\includegraphics[clip,width=7cm]{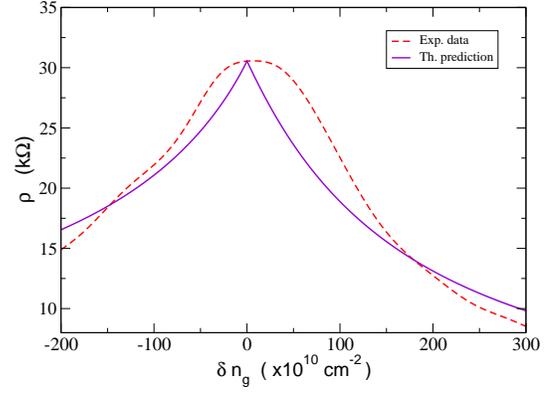}
\par
\end{center}
\caption{Function $[\rho(V_G,c_S,n_i)-\rho(V_G,0,n_i)]/c_S$,
as measured for the Pt-1 sample, expressed in terms of the doping level $\delta n_g$
for the coverage $c_S=0.025$ (red-dashed curved), plotted against the 
inverse of \eqref{eqB16} (purple continuous curve).}
\label{fig:6} 
\end{figure}

\begin{figure}
\begin{center} 
\includegraphics[clip,width=7cm]{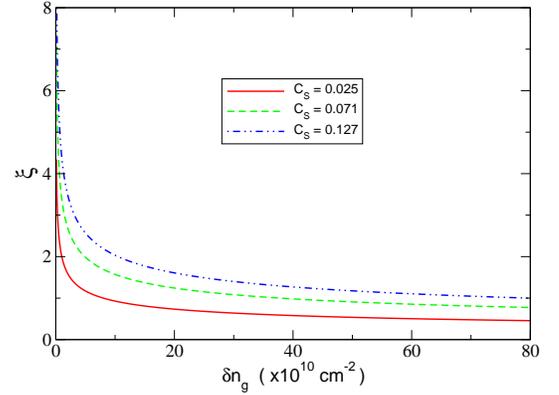}
\par
\end{center}
\caption{Ratio $\xi$, as given by (\ref{eqB9B}), expressed in terms of the 
doping level $\delta n_g$, for sample Pt-1, for the coverages 
$c_S=0.025,\, 0.071$ and $0.127$ ML (rgb).}
\label{fig:7} 
\end{figure}

\begin{figure}
\begin{center} 
\includegraphics[clip,width=7cm]{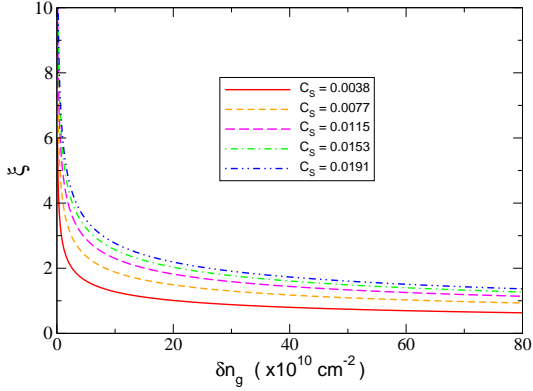}
\par
\end{center}
\caption{Ratio $\xi$, as given by (\ref{eqB9B}), expressed in terms of the 
doping level $\delta n_g$, for sample Ti-1, for the coverages $c_S=0.0038,
0.0077, 0.0115, 0.0153$ and $0.0191$ ML (romgb).}
\label{fig:7Ti} 
\end{figure}

In order to understand the reason for the lack of agreement between the 
above theory and the experiments of \cite{pi}, we consider the behaviour of the ratio $\xi$,
introduced above, for the samples Pt-1 and Ti-1 (the behaviour observed for
the sample Pt-3 is analogous to that of Pt-1). The plots are presented
in figures \ref{fig:7} and \ref{fig:7Ti}.
These plots indicate that, except for the lowest coverages, $\xi>1$ in the 
whole range of doping displayed, and thus that the independent cluster
approximation is unlikely to work for such high
coverages. Note that this is purely a geometric effect, caused by the
small size of the clusters. One can also write $R_c=\sqrt{\frac{n_{amc}A_c}{\pi c_S}}=a\,
\sqrt{\frac{\sqrt{3}n_{amc}}{2\pi c_S}}$, where $a=2.46$ A is the length
of the primitive cell of graphene. For the lowest coverage
studied in the two Pt samples that we analysed, i.e. $c_S=0.0065$ ML for the Pt-3 sample, 
with $n_{amc}=60$ atoms, $R_c\approx 12$ nm. In the case of the Ti-1
sample, $n_{amc}=2$ atoms and for the lowest coverage studied $c_S=0.0038$,
$R_c\approx 3$ nm (for higher coverages, $R_c$ is even smaller). 
Thus, in order to test the theory presented, the experiments performed 
would need to be repeated on samples presenting much lower
concentrations of the deposited TM atoms (less than $1$\% for Pt-covered samples
and less than $0.2$\% for Ti-covered samples). In addition, an appropriate
characterisation of the clusters and their size distribution would
also be required. With regard to the range of coverages studied experimentally
by \cite{pi}, one should also note that if one takes the doping level 
of graphene to be $\delta n_g \approx 10^{12}$ cm$^{-2}$, one has that 
$k_F\approx 10^{6}$ cm$^{-1}$. Since $R_c\approx 10$ nm or less, 
$k_F R_c\sim 1$. This is yet another indication that a theory based
on independent scattering centres is unlikely to work at this range
of coverages. One should not expect that the contribution
of multiple scattering to the resistivity $\rho_{MS}(\delta n_g,c_S,n_i)$
in (\ref{eqB15}) is a small quantity, in particular in the neighbourhood
of the Dirac point\cite{ferreira}. 
 
\section{Conclusions}
\label{secC}
In this paper, we performed a thermodynamic analysis of the problem
of doping of graphene by TM clusters and computed the magnitude of
the chemical interaction necessary to explain the electron doping
of graphene by the transition metals Pt and Ti, the former having
a bulk work function that is more than $1$ eV larger than the
work function of graphene.
We have shown that the enhancement of such interaction with respect
to the case studied in\cite{gio1,gio2} is due to the finite size
of the TM clusters. We have also determined the scattering potential 
induced in a graphene sheet by spherical TM clusters and its 
contribution to the resistivity of the sample in the FBA. We have
shown that regime of coverages for which the transport theory
presented is likely to have a predictive power is below those 
coverages considered in the experiments of Pi et al. and thus one 
would need to repeat  such experiments in these regimes in order to 
fully test such a theory.
\\ \\
\textbf{Acknowledgements:} We acknowledge helpful dis\-cus\-sions
with K. McCreary, R. Kawakami, P. Fulde, A. Ferreira and E. Lage. J.E.S. 
acknowledges support by FCT under Grant No.PTDC/FIS/64404/2006 
and by the Visitors Program of MPIPkS at the different stages of this work.
A.H.C.N. acknowledges support from the DOE grant DE-FG02-08ER46512 and the 
ONR grant MURI N00014-09-1-1063.
\appendix
\section{Calculation of a cluster's ionisation potential based on a thermodynamic argument}
\label{appO}
One can also determine the ionisation potential of a cluster from thermodynamic
considerations. One starts by considering the expression for the energy (\ref{eq01})
in the case of a single cluster ($N_c=1$). Since the expression (\ref{eq01})
is valid for a system that is neutral, the extraction of a charge
$-e$ from the cluster requires this charge to be replaced in the graphene sheet.
As discussed in appendix \ref{secA}, the placement of such a charge in the graphene
sheet can be pictured as the withdrawal of the image charge $+e$ of the
cluster charge $-e$, as the latter one is removed to infinity. Thus, one has from
(\ref{eq01}) that
\begin{eqnarray}
I_S-W_g&=&{\cal E}(N_S-1,N_g+1,N_{Si})-{\cal E}(N_S,N_g,N_{Si})\nonumber\\
&=&-\overline{\mu}_S+\overline{\mu}_g+e(D_S-D_g)\nonumber\\
&&\mbox{}-\frac{e^2}{C_S}\,(N_S-N_S^0)+\frac{e^2}{2C_S}\,.
\label{eqapA1}
\end{eqnarray}
Provided that the area of the graphene sheet is large, $W_g=-\overline{\mu}_g+eD_g$, and thus
\begin{equation}
I_S=-\overline{\mu}_S+eD_S-\frac{e^2}{C_S}\,(N_S-N_S^0)+\frac{e^2}{2C_S}\,.
\label{eqapA2}
\end{equation}
One can consider the difference in the value of the ionisation potential $I_S$
and the ionisation potential $I_S^0$ in a situation where the cluster is placed
very far away from the graphene sheet. One has
\begin{eqnarray}
I_S-I_S^0&\!=\!&-\Delta\overline{\mu}_S+e(D_S-D_S^0)
-e^2\left(\frac{1}{C_S}-\frac{1}{C_S^0}\right)
\Delta N_S\nonumber\\
&&\mbox{}+\frac{e^2}{2}\left(\frac{1}{C_S}-\frac{1}{C_S^0}\right)\,.
\label{eqapA3}
\end{eqnarray}
Note that $D_S^0$ is not identical to $D_{B}$, due to the finite size of the cluster.
In appendix \ref{secA}, we will show how the last two terms of (\ref{eqapA3}) are
obtained using the method of images to compute $I_S$ for the case of a spherical cluster.

Finally, one can use the equilibrium condition (\ref{eq03}) to 
write (\ref{eqapA1}) as
\begin{equation}
I_S=W_g+\frac{e^2}{2C_S}=-\overline{\mu}_g+eD_g+\frac{e^2}{2C_S}\,.
\label{eqapA4}
\end{equation}

As stated above, one can explicitly compute the capacitance of a spherical cluster of
radius $R$, whose centre lies at a distance $L>R$, see appendix \ref{secA}.
In this case, and in the limit $R,L\rightarrow \infty$, with $\eta=R/2L$ finite, i.e.
for graphene adsorbed on the bulk transition metal, as considered
by \cite{gio1,gio2}, $C_S\rightarrow\infty$, and the transition metal 
ionisation potential and the work-function of graphene become equal at 
equilibrium, as one would expect (in this case, $I_S$ would just reduce to the
TM work-function). For a spherical metal cluster of finite
radius, there is an extra contribution to its ionisation potential, coming from the
effect of the image charge, already present in the case of an isolated 
spherical cluster\cite{hyper}. This is the last term of the rhs of equation 
(\ref{eqapA4}).

One can also compute the cluster's electron affinity $A_S$ using the above
argument. In this case, one is withdrawing an electron from an over-charged 
cluster and delivering it to the graphene sheet. One has
\begin{eqnarray}
A_S-W_g&=&{\cal E}(N_S,N_g,N_{Si})-{\cal E}(N_S+1,N_g-1,N_{Si})\nonumber\\
&=&-\overline{\mu}_S+\overline{\mu}_g+e(D_S-D_g)\nonumber\\
&&\mbox{}-\frac{e^2}{C_S}\,(N_S-N_S^0)-\frac{e^2}{2C_S}=-\frac{e^2}{2C_S}\,,
\label{eqapA5}
\end{eqnarray}
where we have again used the equilibrium condition (\ref{eq03}). Note, in
closing, that $W_g=\frac{1}{2}(I_S+A_S)$, as one would expect.

\section{Calculation of the electrostatic contribution to the ionisation potential of
a spherical cluster close to a grounded plane and of the electrostatic
potential created by it}
\label{secA}

We will now compute the ionisation potential of a system composed by a single 
metallic spherical cluster and a grounded graphene plane using the method of
images, by considering the electrostatic work necessary to extract a single
electron from the cluster. These considerations will also allow us to write
an explicit expression for the system's capacitance and for the electrostatic 
potential created by the cluster in the upper-half space, i.e. the quantity 
$v(\vecg{r})$, introduced in section \ref{secB}, necessary to determine the 
scattering potential of carriers in graphene due to the presence of the spherical cluster.

As above, the sphere contains a total charge $Q_S$ and we will extract
a charge $q_0$ from it, leaving a total charge $Q_S-q_0$ in it.
The charge $q_0$ is the elementary charge $-e$. The ionisation potential 
of this system is the energy necessary to displace $q_0$ from a distance 
$d$ away from the surface of the sphere to infinity. The distance $d$ at 
which one begins to perform work to extract the
charge $q_0$ is a regularisation parameter necessary to take into account the
singular nature of the Coulomb interaction, but one can also interpret it
physically as being the distance beyond which quantum corrections to the
Coulomb law become negligible.

The conditions of the problem are as described in section \ref{secB}. In order 
to determine the force on $q_0$ as it is displaced from $z=d+R$ to $z=\infty$, 
one needs to determine the potential created by the presence of the sphere 
and of the plane at the position of the charge. Such potential can be determined 
by the method of images, as shown below.

In the absence of a conducting plane, the solution of the problem is trivial. 
If $q_0$ is located at $z=z_0$, one places an image charge of magnitude $q_1=-q_0 R/z_0$ 
in the interior of the sphere at $z_1=R^2/z_0$ and an image charge $q_2=Q_S-q_0-q_1$ 
at the centre of the sphere. These three charges guaranty that the surface of 
the sphere is equipotential and that the sphere has an overall charge
$Q_S-q_0$ in it. In the presence of a grounded plane, these three charges no 
longer guaranty that the surface $z=-L$ is an equipotential. We therefore 
take $q_2$ to have an arbitrary value for the moment and
consider three image charges $q_3=-q_0$ at $z_3=-(2L+z_0)$, $q_4=-q_1$ at 
$z_4=-(2L+z_1)$ and $q_5=-q_2$ at $z_5=-2L$, located below the plane. 
These three charges will guaranty that the plane $z=-L$ is 
an equipotential. However, the surface of the sphere is no 
longer an equipotential. We therefore place three 
image charges in the interior of the sphere,
$q_6=-q_3 \frac{R}{\mid z_3\mid}=
\frac{q_0}{2L+z_0}$ at $z_6=-\frac{R^2}{\mid z_3\mid}=-\frac{R^2}{2L+z_0}$,
$q_7=-q_4 \frac{R}{\mid z_4\mid}=q_1 \frac{R}{2L+z_1}$ at $z_7=-\frac{R^2}{\mid z_4\mid}
=-\frac{R^2}{2L+z_1}$ and $q_8=-q_5 \frac{R}{\mid z_5\mid}
=q_2 \frac{R}{2L}$ at $z_8=-\frac{R^2}{\mid z_5\mid}=-\frac{R^2}{2L}$. In order to 
balance the potential at the surface of the plane, we now need to place three 
image charges below the plane, followed by three image charges 
inside the sphere and so on {\it ad infinitum}. 
The arguments above suggest that the recurrence relation
between the charges inside the sphere is given by
\begin{equation}
\label{eq1}
\left\{
\mbox{
\begin{tabular}{c}   
$q_{6(n+1)+\alpha}=q_{6n+\alpha}\,\frac{R}{2L+z_{6n+\alpha}}$\\
$z_{6(n+1)+\alpha}=-\frac{R^2}{2L+z_{6n+\alpha}}$
\end{tabular}
}\right.
\end{equation}
where $n\geq 0$ and $\alpha=0,1,2$. The recurrence relation between the charges 
located inside the sphere and below the plane is simpler, 
$q_{6n+3+\alpha}=-q_{6n+\alpha}$, $z_{6n+3+\alpha}=-(2L+z_{6n+\alpha})$.

The charge on the sphere is given by 
\begin{eqnarray}
\label{eq2}
Q_S&=&q_0+q_1+q_2+q_6+q_7+q_8+q_{12}+q_{13}+q_{14}+\cdots\nonumber\\
&=&\sum_{n=0,\alpha}^\infty
\,q_{6n+\alpha}\,.
\end{eqnarray}
This equation determines the charge $q_2$ in terms of $Q_S$ and $q_0$. Let us rewrite
the recursion relation above in a slightly different form. We define $\pi^\alpha_n=q_{6n+\alpha}$,
$w_{n}^\alpha=\frac{z_{6n+\alpha}}{2L}$ and $\eta=\frac{R}{2L}<\frac{1}{2}$. We have
\begin{equation}
\label{eq3}
\left\{
\mbox{
\begin{tabular}{c}   
$\pi_{n+1}^\alpha=\pi_{n}^\alpha\,\frac{\eta}{1+w_n^\alpha}$\\
$w_{n+1}^\alpha=-\frac{\eta^2}{1+w_n^\alpha}$
\end{tabular}
}\right.
\end{equation}
We now define $u_n^\alpha=\frac{1}{\pi_n^\alpha}$ \cite{jeans}. It is easy to
see from (\ref{eq3}) that we have $u_{n+1}^\alpha=u_{n}^\alpha\,
\frac{1+w_n^\alpha}{\eta}$ and $u_{n-1}^\alpha=u_{n}^\alpha\,
\frac{\eta}{1+w_{n-1}^\alpha}$. Adding the two equations and using 
the recursion relation for $w_{n}^\alpha$, we have
\begin{equation}
\label{eq4}
u_{n+1}^\alpha+u_{n-1}^\alpha=\frac{u_n^\alpha}{\eta}\,.
\end{equation}
This is a linear recursion relation, with solution $u_n^\alpha=A_+^\alpha\,s_+^n+A_-^\alpha\,s_-^n$,
where $s_\pm$ are the solutions of the quadratic equation $\eta(s_\pm^2+1)-s_\pm=0$, which 
are given by $s_\pm=\frac{1}{2\eta}\pm\frac{\sqrt{1-4\eta^2}}{2\eta}$, with $s_+s_-=1$ and $
s_+>1, s_-<1$. For definitiveness, we will call $s_-=\lambda, s_+=\lambda^{-1}$. In this case, we have
\begin{equation}
\label{eq5}
\left\{
\mbox{
\begin{tabular}{c}   
$u_n^\alpha=A_+^\alpha\,\lambda^{-n}+A_-^\alpha\,\lambda^n$\\
$w_n^\alpha=-\frac{\eta(A_+^\alpha\,\lambda^{1-n}+A_-^\alpha\,\lambda^{n-1})}{
A_+^\alpha\,\lambda^{-n}+A_-^\alpha\,\lambda^n}$
\end{tabular}
}\right.,
\end{equation}
where we have made use of the relation $\eta(\lambda^{-1}+\lambda)=1$.
The values of the coefficients $A_{\pm}^\alpha$ are determined from the 
known values
of $u_0^\alpha, w_0^\alpha$, namely $u_0^0=q_0^{-1}$, $w_0^0=w_0=z_0/2L$, 
$u_0^1=-q_0^{-1}\frac{w_0}{\eta}$, $w_0^1=\frac{\eta^2}{w_0}$ and 
$u_0^2=q_2^{-1}$, $w_0^2=0$. One obtains $A_+^0=\frac{1}{q_0\eta}\,\frac{w_0+
\eta\lambda^{-1}}{\lambda^{-1}-\lambda}$, $A_-^0=-\frac{1}{q_0\eta}\,
\frac{w_0+\eta\lambda}{\lambda^{-1}-\lambda}$, $A_+^1=-\frac{1}{q_0\eta\lambda}\,\frac{w_0+
\eta\lambda}{\lambda^{-1}-\lambda}$, $A_-^1=\frac{\lambda}{q_0\eta}\,
\frac{w_0+\eta\lambda^{-1}}{\lambda^{-1}-\lambda}$ and $A_+^2=\frac{1}{q_2}\,
\frac{\lambda^{-1}}{\lambda^{-1}-\lambda}$, $A_-^2=-\frac{1}{q_2}\,
\frac{\lambda}{\lambda^{-1}-\lambda}$. Substituting these relations and
equation (\ref{eq5}) in equation (\ref{eq2}), one obtains for $q_2$ the following result
\begin{eqnarray}
\label{eq6}
q_2(\lambda,\xi)&\!=\!&\frac{Q_S-q_0}{\lambda^{-1}-\lambda}
g^{-1}(\lambda,1)-\frac{q_0(1-\xi)}{\lambda^{-1}-\lambda}\nonumber\\
&&\mbox{}\times\left(g(\lambda,\xi)
-\xi^{-1}g(\lambda,\xi^{-1})\right)g^{-1}(\lambda,1),
\end{eqnarray}
where $\xi=\frac{w_0+\eta\lambda}{w_0+\eta\lambda^{-1}}<1$ and \cite{Note3}
\begin{equation}
\label{eq6A}
g(\lambda,\xi)=\sum_{n=1}^{\infty}\,
\frac{\lambda^n}{1-\xi\lambda^{2n}}\,.
\end{equation}

Note that one can use this result to compute 
the capacitance of the system composed by the sphere and the plane. For such a calculation, 
one takes $q_0=0$ (or $w_0\rightarrow \eta$, i.e. $\xi\rightarrow \lambda$). 
The charge of the sphere is $Q_S$ and the potential at its 
surface is given by $V_S=\frac{q_2}{4\pi\epsilon_0 R}$. 
Hence, the capacitance $C_S=Q_S/V_S$ is, using
(\ref{eq6}) with $q_0=0$, given by 
\begin{equation}
\label{eq6B}
C_S=4\pi\epsilon_0 R\,(1-\lambda^2)\,g(\lambda,1)/\lambda\,.
\end{equation}
In terms of $\eta$, the first few terms of this series are $C_S=4\pi\epsilon_0 R\left(\,1+\eta+
\frac{\eta^2}{1-\eta^2}+\cdots\,\right)$.

The potential created by the image charges along the $z$-axis ($z\geq R$) is given by
\begin{eqnarray}
\label{eq7}
V_{im}(z,z_0)&=&\frac{1}{4\pi\epsilon_0}\,\left\{\sum_{[n,\alpha]}^\infty\,
\frac{q_{6n+\alpha}}{z-z_{6n+\alpha}}\right.\nonumber\\
&&\left.-\sum_{n,\alpha}^\infty\,
\frac{q_{6n+\alpha}}{z+2L+z_{6n+\alpha}}\right\}\,,
\end{eqnarray}
where we have used the recursion relation between the image charges on the plane and those on the sphere, and where the notation $[n,\alpha]$ indicates
that the term $n=0, \alpha=0$ (potential created by $q_0$) is absent from the first sum. 
The notation $V_{im}(z,z_0)$ indicates that the potential depends on $z_0$ through its dependence on the position and magnitude
of the image charges. Expressing $q_{6n+\alpha}$ in terms of $u_n^\alpha$ and $z_{6n+\alpha}$
in terms of $w_n^\alpha$ and using the rescaled variable $w=z/2L$, we have that $V_{im}(w,w_0)$ is given by
\begin{eqnarray}
\label{eq8}
V_{im}(w,w_0)&=&\frac{1}{8\pi\epsilon_0
  L}\,\left\{\sum_{[n,\alpha]}^\infty\,\frac{1}{u_n^\alpha(w-w_n^\alpha)}\right.\nonumber\\
&&\left.-\sum_{n,\alpha}^\infty\,\frac{1}{u_n^\alpha(w+1+w_n^{\alpha})}\right\}\,.
\end{eqnarray}
Substituting the recursion relations given by (\ref{eq5}) above, we obtain
for ${\cal U}_{im}(w,w_0)=8\pi\epsilon_0 L\,V_{im}(w,w_0)$
\begin{eqnarray}
\label{eq9}
{\cal U}_{im}(w,w_0)&\!\!=\!\!&\sum_{[n,\alpha]}^\infty\frac{1}{A_+^\alpha\lambda^{-n}(w+\eta\lambda)+A_-^\alpha\lambda^n(w+\eta\lambda^{-1})}\nonumber\\
&&\mbox{}\!\!-\sum_{n,\alpha}^\infty\frac{1}{A_+^\alpha\lambda^{-n}(w+\eta\lambda^{-1})+A_-^\alpha\lambda^n(w+\eta\lambda)}.
\nonumber\\
&&
\end{eqnarray}

The potential $v(\vecg{r})$ created by the cluster on the subspace $z\geq -L$,
when its charge is equal to $Q_S$, which was introduced in section \ref{secB}, 
can also be computed in a manner analogous to $V_{im}(z,z_0)$. In this case 
one sets $q_0=0$, as in the calculation of the capacitance, in the recursion relation (\ref{eq5}).
One is now interested in the dependence of $v(\vecg{r})$ both on 
$z$ and on the radial coordinate $r$ along the $xy$ plane.

This potential is given by
\begin{eqnarray}
\label{eqB2}
v(\vecg{r})&=&\frac{1}{4\pi\epsilon_0}\,\sum_{n=0}^\infty\,\left(\,
\frac{q_{6n+2}}{[r^2+(z-z_{6n+2})^2]^{1/2}}\right.\nonumber\\
&&\left.-\frac{q_{6n+2}}{[r^2+(z+2L+z_{6n+2})^2]^{1/2}}\,\right)\,,
\end{eqnarray}
where $r=\sqrt{x^2+y^2}$ is the distance in the graphene sheet
and where $q_{6n+2}$ and $z_{6n+2}$ are given by
\begin{equation}
\label{eqB2A}
\left\{
\mbox{
\begin{tabular}{c}   
$q_{6n+2}=\frac{Q_S\,g^{-1}(\lambda,1)}{\lambda^{-(n+1)}-\lambda^{(n+1)}}$\\
$z_{6n+2}=-R\,\frac{\lambda^{-n}-\lambda^{n}}{\lambda^{-(n+1)}-\lambda^{(n+1)}}$
\end{tabular}
}\right..
\end{equation}

One can now use (\ref{eq9}) to compute the ionisation potential of the system cluster-graphene plane.
This quantity is equal to the work of the external force necessary to
transport the charge $q_0$ from $z=R+d$ up to $z=\infty$ quasi-statically,
i.e. $I_S(d)=\int_{R+d}^\infty\,dz_0\,F_{ext}(z_0)$, where
$F_{ext}(z_0)=\frac{q_0}{16\pi\epsilon_0L^2}\,\frac{\partial {\cal U}_{im}}{\partial w}\mid_{w=w_0}$. 

Substituting the values $A_{\pm}^\alpha$ obtained above in (\ref{eq9}) and
performing the derivative of ${\cal U}_{im}(w,w_0)$ with respect to $w$ at $w=w_0$, one
obtains the rather lengthy expression for $f_{ext}(w_0)=\frac{1}{\eta(\lambda^{-1}-\lambda)}\frac{\partial {\cal U}_{im}}{\partial w}\mid_{w=w_0}$,
\begin{eqnarray}
\label{eq10}
f_{ext}(w_0)&\!\!=\!\!&q_0\,\sum_{n=0}^\infty\,\frac{(w_0+\eta\lambda^{-1})\lambda^{-n}-(w_0+\eta\lambda)\lambda^n}{[\,(w_0+\eta\lambda^{-1})^2\lambda^{-n}
-(w_0+\eta\lambda)^2\lambda^n\,]^2}\nonumber\\
&&\mbox{}+q_0\,\sum_{n=1}^\infty\,\frac{(w_0+\eta\lambda)\lambda^{-n}-(w_0+\eta\lambda^{-1})\lambda^n}{[\,(w_0+\eta\lambda)^2\lambda^{-n}
-(w_0+\eta\lambda^{-1})^2\lambda^n\,]^2}\nonumber\\
&&\mbox{}-q_0\,\sum_{n=1}^\infty\,\frac{(w_0+\eta\lambda^{-1})\lambda^{-n}-(w_0+\eta\lambda)\lambda^n}{[\,(w_0+\eta\lambda^{-1})\,(w_0+\eta\lambda)\,(\lambda^{-n}-\lambda^n)\,]^2}\nonumber\\
&&\mbox{}-q_0\,\sum_{n=1}^\infty\,\frac{(w_0+\eta\lambda)\lambda^{-n}-(w_0+\eta\lambda^{-1})\lambda^n}{[\,(w_0+\eta\lambda^{-1})\,(w_0+\eta\lambda)\,(\lambda^{-n}-\lambda^n)\,]^2}\nonumber\\
&&\mbox{}+q_2\,\sum_{n=1}^\infty\,\frac{\lambda^{-n}-\lambda^n}{[\,(w_0+\eta\lambda^{-1})\lambda^{-n}
-(w_0+\eta\lambda)\lambda^n\,]^2}\nonumber\\
&&\mbox{}-q_2\,\sum_{n=1}^\infty\,\frac{\lambda^{-n}-\lambda^n}{[\,(w_0+\eta\lambda)\lambda^{-n}
-(w_0+\eta\lambda^{-1})\lambda^n\,]^2}\,.\nonumber\\
&&
\end{eqnarray}
This expression has to be integrated so as to obtain the ionisation potential
$I_S(d)=\frac{q_0\sqrt{L^2-R^2}}{8\pi\epsilon_0 L^2}\,\int_{\eta+\gamma}^\infty\,dw_0\,f_{ext}(w_0)$, where $\gamma=d/2L$. It is more convenient 
to express this integral in terms of the variable $\xi$ introduced above. The result is
\begin{eqnarray}
\label{eq11}
I_S(d)&=&\frac{q_0^2}{8\pi\epsilon_0
  \sqrt{L^2-R^2}}\,\int_{1/\delta}^1\,d\xi\,(1-\xi)\nonumber\\
&&\mbox{}\times\left[\,\sum_{n=0}^\infty\frac{\lambda^n\,(1-\xi\lambda^{2n})}{(1-\xi^2\lambda^{2n})^2}\right.\nonumber\\
&&\left.+\frac{1}{\xi^3}\,\sum_{n=1}^\infty\frac{\lambda^n\,(1-\xi^{-1}\lambda^{2n})}{(1-\xi^{-2}\lambda^{2n})^2}\,\right]\nonumber\\
&&\mbox{}+\frac{q_0^2\,g(\lambda,1)}{8\pi\epsilon_0 \sqrt{L^2-R^2}}\,\int_{1/\delta}^1\,d\xi\,\left(1-\frac{1}{\xi^2}\right)\,\nonumber\\
&&\mbox{}+\frac{q_0}{4\pi\epsilon_0
  R}\,\int_{1/\delta}^1\,d\xi\,q_2(\lambda,\xi)\,\left[\,\sum_{n=1}^\infty\frac{\lambda^n\,(1-\lambda^{2n})}{(1-\xi\lambda^{2n})^2}\right.\nonumber\\
&&\left.-\frac{1}{\xi^2}\,
\sum_{n=1}^\infty\frac{\lambda^n\,(1-\lambda^{2n})}{(1-\xi^{-1}\lambda^{2n})^2}\,\right]
\,,
\end{eqnarray}
where $q_2(\lambda,\xi)$ is given in equation (\ref{eq6}), where we have also introduced $g(\lambda,\xi)$ and where $\delta=\frac{\eta+\gamma+\eta\lambda^{-1}}{\eta+\gamma+\eta\lambda}>1$
and $\delta<\lambda^{-1}$, being equal to it in when $d=0$.
The first term in this expression can be integrated using the following identity
\begin{eqnarray}
\label{eq11a}
\frac{1}{2(1-\xi)}\frac{\partial}{\partial \xi}\left\{(1-\xi)^2[g(\lambda,\xi^2)+
\xi^{-2}g(\lambda,\xi^{-2})]\right\}=\\
-\left[\sum_{n=1}^\infty\frac{\lambda^n(1-\xi\lambda^{2n})}{(1-\xi^2\lambda^{2n})^2}+\frac{1}{\xi^3}
\sum_{n=1}^\infty\frac{\lambda^n(1-\xi^{-1}\lambda^{2n})}{(1-\xi^{-2}\lambda^{2n})^2}\right].\nonumber
\end{eqnarray}
The second term involves a trivial integral. As for the third term, it can also be integrated
if one notes that the following identity holds
\begin{eqnarray}
\label{eq12}
-\frac{\partial}{\partial
  \xi}\left\{\,(1-\xi)[\,g(\lambda,\xi)-\xi^{-1}\,g(\lambda,\xi^{-1})\,]\,\right\}=\\
\sum_{n=1}^\infty\frac{\lambda^n\,(1-\lambda^{2n})}{(1-\xi\lambda^{2n})^2}-\frac{1}{\xi^2}\,
\sum_{n=1}^\infty\frac{\lambda^n\,(1-\lambda^{2n})}{(1-\xi^{-1}\lambda^{2n})^2}\,.\nonumber
\end{eqnarray}
Substituting these results and (\ref{eq6}) in (\ref{eq11}), performing the resulting integrals and putting $q_0=-e$, one obtains after some trivial manipulations, the final result
\begin{eqnarray}
\label{eq13}
I_S(d)&=&\frac{e^2}{16\pi\epsilon_0(L+R+d)}\nonumber\\
&&\mbox{}+\frac{e^2\,\sqrt{L^2-R^2}}{4\pi\epsilon_0
  [\,L+R+d-\sqrt{L^2-R^2}\,]^2}\nonumber\\
&&\mbox{}\times[\,g(\lambda,\delta^2)+\delta^{-2}\,g(\lambda,\delta^{-2})\,]
\nonumber\\
&&\mbox{}-\frac{e^2\,\sqrt{L^2-R^2}\,g(\lambda,1)}{2\pi\epsilon_0\,[\,d^2+2(R+L)(R+d)\,]}\nonumber\\
&&\mbox{}+\frac{(e^2+e\,Q_S)\,g^{-1}(\lambda,1)\,}{4\pi\epsilon_0\,[\,L+R+d-\sqrt{L^2-R^2}\,]}\nonumber\\
&&\mbox{}\times[\,g(\lambda,\delta)-\delta^{-1}\,
g(\lambda,\delta^{-1})\,]\nonumber\\
&&\mbox{}-\frac{e^2\,\sqrt{L^2-R^2}\,g^{-1}(\lambda,1)\,}{4\pi\epsilon_0\,[\,L+R+d-\sqrt{L^2-R^2}\,]^2}
\nonumber\\
&&\mbox{}\times[\,g(\lambda,\delta)-\delta^{-1}\,
g(\lambda,\delta^{-1})\,]^2\,.
\end{eqnarray}

In the limit $L\rightarrow \infty$, we are left with an isolated charged cluster.
We obtain from (\ref{eq13}) the known result
\begin{eqnarray}
\label{eq14}
I_S^0(d)&=&\frac{e^2}{4\pi\epsilon_0}\left(\,
\frac{R}{2d(d+2R)}+\frac{1}{R+d}-\frac{R}{2(R+d)^2}\,\right)\nonumber\\
&&\mbox{}+\frac{e\,Q_S}{4\pi\epsilon_0(R+d)}\,.
\end{eqnarray}
The only term in (\ref{eq13}) that is singular in the limit $d\rightarrow 0$ is
the first term of the series $g(\lambda,\delta^2)$ ($\delta\rightarrow \lambda^{-1}$
in this limit). Moreover, for finite $d$, this term is equal to the
corresponding singular term in (\ref{eq14}). Thus, $\Phi(d)=I_S(d)-I_S^0(d)$ 
is a regular function in the limit $d\rightarrow 0$. It equals
\begin{equation}
\Phi(0)=e\,Q_S\left(\frac{1}{C_S}-\frac{1}{C_S^0}\right)
+\frac{e^2}{2}\left(\frac{1}{C_S}-\frac{1}{C_S^0}\right)\,,
\label{eq15}
\end{equation}
with $C_S^0=4\pi\epsilon_0 R$ and where $C_S$ is given by (\ref{eq6B}).
Since $Q_S=-e\,\Delta N_S$, we see that the result (\ref{eq15}) 
is equal to the last two terms of (\ref{eqapA2}), which correspond to
the electrostatic contribution to $I_S$, the only one considered here.
This justifies the statement made in appendix \ref{appO}.


\begin{thebibliography}{100}
\bibitem{Tichan} According to the studies of \cite{chan,gio2}, Ti is chemically
  adsorbed in graphene and does not preserve the Dirac point in the newly
computed band structure. However, since the maximum concentration of Ti studied 
by \cite{pi} is below $1\%$, one does not expect that in this case the band
structure will be significantly changed with respect to pristine graphene.
\bibitem{NoteCC} The assumption that the cross-capacitance between two
clusters $S$ and $S'$ is solely due to the presence of the graphene plane
can be expressed mathematically as $C_{SS'}^{-1}=2C_{Sg}^{-1}-C_{gg}^{-1}$, where
$C_{SS'}^{-1}$, $C_{Sg}^{-1}$ and $C_{gg}^{-1}$ are, respectively, the inverse
cross capacitance of two clusters, the inverse cross capacitance between
a cluster and the graphene plane and the inverse self-capacitance of graphene.
Writing the joint capacitance of two clusters as $\tilde{C}_S^{-1}=
C_{SS}^{-1}+C_{S'S'}^{-1}-2C_{SS'}^{-1}=2(C_{SS}^{-1}-C_{SS'}^{-1})$, 
since $C_{S'S'}^{-1}=C_{SS}^{-1}$, and using the equation above, 
we see that this expression is equivalent to
the equality $\tilde{C}_S^{-1}=2C_S^{-1}$, where 
$C_S^{-1}=C_{SS}^{-1}+C_{gg}^{-1}-2C_{Sg}^{-1}$ is the joint capacitance
of a cluster and the graphene plane. Note that in the derivation of
(\ref{eq01}) the assumption that $C_{S Si}^{-1}= C_{Sg}^{-1}+C_{gSi}^{-1}-C_{gg}^{-1}$,
i.e. that the cross-capacitance between the clusters and the Si sheet
is due to two capacitors in series, is also used.
\bibitem{noteccl} Equation (\ref{eq023}) is also valid for a simple capacitor
geometry in the absence of deposited metal. In that case, one sees that for
the parameters as given in experiments made with capacitor
like devices, $\mid \Omega\mid\gg \Lambda^2/4$ and hence the doping of graphene
is very well described by a simple capacitor law \cite{novo1}, except in the
neighbourhood of the Dirac point.
\bibitem{radiusTi} A spherical cluster of Ti atoms with $R=1.88$ A contains
  only 2 atoms of titanium. It is thus debatable if a continuum
theory such as the one presented above is applicable. The relative agreement
obtained may be accidental, see \cite{ioeTi}.
\bibitem{LmR} The experimental results of \cite{pi} allow one to
place a limit on the minimal distance between the clusters and the graphene
sheet, i.e. $L-R\leq 3$ A. We have therefore chosen $L-R=2.5$ A, although
the results for $\Delta_c$ are only very weakly dependent on $L-R$, unlike
what happens with $R$.
\bibitem{NoteFS} The result of references\cite{gio1,gio2} for the induced
  dipoles was obtained for a system of a single graphene sheet adsorbed on
  a bulk transition metal and hence does not take into consideration the effect 
  of the finite size of the clusters.
\bibitem{ioeTi} The ionisation potential of a Ti cluster computed from
(\ref{eq024}) is $\approx 7.5$ eV. This value is similar to the average
ionisation energy of $8.0$ eV of an isolated Ti atom, if we consider 
the average doping of $-0.174$ e/Ti atom, using the first and second ionisation energies of the titanium atom.
The relative coincidence of these two values may help explain why one is still able to apply
a continuum theory to describe essentially isolated titanium atoms.
\bibitem{objgio} The computed work-function of graphene adsorbed on (111)
platinum in \cite{gio1,gio2} is $4.87$ eV with $c_S=0.75$. 
This is to be compared with the value of the work function at zero gate 
voltage obtained from (\ref{eq024}) for the largest coverage $c_S=0.171$
considered above, if we neglect the induced dipole, which is equal to $4.33$
eV. Hence, $e\Delta D_g>0.55$ eV, at large coverages.
\end{thebibliography}
\end{document}